\documentclass[
aps,%
12pt,%
final,%
notitlepage,%
oneside,%
onecolumn,%
nobibnotes,%
nofootinbib,%
superscriptaddress,%
noshowpacs,%
centertags]%
{revtex4}
\usepackage{bm}
\usepackage{color}
\newlength{\lenl}
\newlength{\lenc}
\newlength{\lenr}
\newcommand{\be}{\begin{equation}}
\newcommand{\ee}{\end{equation}}
\newcommand{\bea}{\begin{eqnarray}}
\newcommand{\eea}{\end{eqnarray}}
\begin{document}
\selectlanguage{english}

\title{Bound States in Gauge Theories as the Poincar\'e Group Representations
%Pion catalysis of the anomalous para-positronium creation
}
\author{A.Yu.~Cherny}
\affiliation{Bogoliubov Laboratory of Theoretical Physics,
Joint Institute for Nuclear Research, Dubna, Russia}
\author{A.E.~Dorokhov}
\affiliation{Bogoliubov Laboratory of Theoretical Physics,
Joint Institute for Nuclear Research, Dubna, Russia}
\author {Nguyen~Suan~Han}
\affiliation{
Department of Theoretical Physics, Vietnam National University}
\author{V.N.~Pervushin}
\email{pervush@theor.jinr.ru}
\affiliation{Bogoliubov Laboratory of Theoretical Physics,
Joint Institute for Nuclear Research, Dubna, Russia}
\author{V.I.~Shilin}
\affiliation{Bogoliubov Laboratory of Theoretical Physics,
Joint Institute for Nuclear Research, Dubna, Russia}
\affiliation{Moscow Institute of Physics and Technology,  Dolgoprudnyy, Russia}

\begin{abstract}

The bound state generating functional is constructed in gauge theories.
 This
construction is based on  the Dirac Hamiltonian approach to gauge theories,
 the Poincar\'e group classification of fields  and their nonlocal bound states,
and the Markov-Yukawa  constraint of irreducibility.
The generating functional contains  additional  anomalous
creations of pseudoscalar bound
states: para-positronium in QED and mesons in QCD in the two gamma processes of the type of
 $\gamma +
\gamma = \pi_0+\text{para-positronium}$.
The functional allows us to
establish physically clear
and transparent relations between the perturbative QCD to its nonperturbative low energy model by means of
normal ordering and the quark and gluon condensates.
%the low-energy QCD  and its  Numbu-Jona-Lasinio
%model with the non-local pion wave function.connect
%Within the obtained bound state
%generalization of Faddeev-Popov (FP) integral, we
   In the limit of small
current quark masses, the Gell-Mann-Oakes-Renner relation is derived from the
Schwinger-Dyson (SD) and Bethe-Salpeter (BS) equations. The constituent quark
masses can be calculated from a self-consistent non-linear equation. %We
%discuss a possible detection of the effect  of  anomalous para-positronium
%creation in the two gamma processes inspired by a pion of the type
%As was shown in the ladder
%approximation via the Hubbard -- Stratanovich representation of fermions, this
%generating functional contains the anomalous creation .

\end{abstract}

%\begin{keyword}

%Pion catalysis of the anomalous para-positronium creation
%Faddeev-Popov integral\sep instantaneous bound states \sep bilocal fields

\pacs{11.55.Hx,
13.60.Hb,
25.20.Lj}
%\end{keyword}

%\pacs{11.55.Hx, 13.60.Hb, 25.20.Lj}
%
%\keywords{Faddeev-Popov integral, instantaneous bound states, Markov-Yukawa
%bilocal fields}

\maketitle

Is dedicated to the 60-th anniversary of the birth of Professor S.I. Vinitsky

%\tableofcontents

\section{Introduction}

At the beginning of the sixties of the twentieth century Feynman found that
the naive generalization of his method of  construction of QED fails in the
non-Abelian theories. The unitary S-matrix in the non-Abelian theory was
obtained in the form of the FP path integral by the brilliant application of
the theory of connections in vector bundle \cite{fp1}. Many physicists are of
opinion that the FP path integral is the highest level of quantum description
of the gauge constrained systems. Anyway, the FP integral at least allows us
to prove the both renormalizability of the unified theory of electroweak
interactions and asymptotic freedom of the non-Abelian theory. However, the
generalization of the FP path integral to the bound states in the non-Abelian
theories still remains a serious and challenging problem.

The bound states in gauge theories are usually considered in the framework of
representations of the homogeneous Lorentz group and the FP functional in one
of the Lorentz-invariant gauges. In particular, the ``Lorentz gauge
formulation'' was discussed in the review \cite{Hay-91} with almost 400 papers
before 1992 on this subject being cited. Presently, the situation is not
changed, because the gauge-invariance of the FP path integral is proved only
for the scattering processes of elementary particles on their mass shells in
the framework of the ``Lorentz gauge formulation'' \cite{f}.

In this paper, we suggest a systematic scheme of the bound state
generalization of FP-functional and S-matrix elements. The scheme is based on irreducible
representations of the nonhomogeneous Poincar\'e group in concordance with
the first QED description of bound states \cite{a15,Salpeter}. This approach includes
the following elements.

\emph{i)} The concept of the in- and out- state rays \cite{logunov} as the
products of the Poincar\'e representations of the Markov-Yukawa bound states
\cite{MarkovYukawa,kad,LukierskiOziewicz,Pervu2}.

\emph{ii)} The split of the potential components from the radiation ones in a
rest frame.

\emph{iii)} Construction of the bound state functional in the presence of the
radiation components. This functional contains the triangle axial anomalies
with additional time derivatives of the radiation components.

\emph{iv)} The joint Hamiltonian approach to the sum of the both standard time
derivatives and triangle anomaly derivatives.

All these elements together leads to new anomalous processes in the strong magnetic fields.
One of them is the two gamma para-positronium creation
 accompanied by the pion
 creation
 of the type of $\gamma+\gamma=\pi_0+$ para-positronium.

 Within the
bound state generalization of the FP integral, we establish physically clear
and transparent relations between the parton  QCD model and the
Numbu-Jona-Lasinio ones \cite{a1,a2,a3}. Below we show that it can be done by
means of the gluon and quark condensates, introduced via the normal ordering.

The paper is organized as follows. Sec.~\ref{sect_R} regards the Poincar\'e
classification of in- and out- states.
In Sec.~3, %\ref{sect_A},
the Dirac method
 of gauge invariant separation of potential and radiation
variables  is considered within QED.  Sec.~4 %\ref{sect_P}
 is devoted to
the bound state generalization of the Faddeev-Popov generating functional.
In Sec.~5,
 %\ref{sect_B},
we discuss the
bound state functional in the presence of the radiation components, which
contains the triangle axial anomalies with additional time derivatives of
these radiation components. Sec.~6 %\ref{sect_H}
is devoted to the axial
anomalies in the NJL model inspired by QCD.
In Appendix A, the bound state functional
in the ladder approximation is considered. In Appendix B, the BS equations are
written down explicitly and discussed.

\section{ Bogoliubov-Logunov-Todorov Rays as In-, Out-States}
\label{sect_R}

According to the general principles of quantum field theory (QFT), physical
states  of the lowest order of perturbation theory are completely covered  by
local fields as  particle-like representations of the Poincar\'e group of
transformations of four-dimensional space-time.

The existence of each elementary particle is associated with a quantum fields
$\psi$. These fields are operators defined in all space-time and acting on
states $|{\cal P},s\rangle$ in the Hilbert space with positively defined
scalar product. The states correspond to the wave functions
$\Psi_\alpha(x)=\langle 0|\psi_\alpha(x)|{\cal P},s\rangle $ of free
particles.

Its algebra is formed by generators of the four translations $\hat {\cal
P}_\mu=i\partial_\mu$ and six rotations $\hat{M}^{\mu\nu} =
i[x_\mu\partial_\nu-x_\nu \partial_\mu]$. The unitary and irreducible
representations are eigen-states of the Casimir operators of mass and spin,
given by
 \bea \label{3-1} \hat{\cal P}^2|{\cal P},s\rangle
&=&m_{\psi}^2~|{\cal P},s\rangle ,\\ \label{3-1w}
 -\hat w_p^2 |{\cal P},s\rangle &=&s(s+1)|{\cal P},s\rangle ,\\ \label{3-1ww}
 \hat w_\rho&=& \frac{1}{2}
 \varepsilon_{\lambda\mu\nu\rho} \hat{\cal P}^\lambda \hat{M}^{\mu\nu}.
 \eea

The unitary irreducible Poincar\'e representations describe wave-like
dynamical local excitations of two transverse   photons
 \bea \label{3-1ph} &&
 A^{T}_{(b)}(t,\textbf{x})=\int
 \frac{d^3k}{(2\pi)^3}\sum\limits_{\alpha=1,2}\frac{1}{\sqrt{2\omega({\bf k})}}
 \varepsilon_{(b)\alpha}\left[
 e^{i
 (\omega_{\textbf{k}}t-\textbf{\textbf{k}\textbf{x}})} A^{+}_{\textbf{k},\alpha}+ e^{-i
 (\omega_{\textbf{k}}t-\textbf{\textbf{k}\textbf{x}})}A^{-}_{\textbf{k},\alpha} \right].
  \eea
  Two independent polarizations
$\varepsilon_{(b)\alpha}$ are perpendicular to the wave vector and to each
other, and the photon dispersion is given by
$\omega_{\textbf{k}}=\sqrt{{\textbf{k}}^2}$.

The creation and annihilation operators of photon obey the commutation
relations $[A^{-}_{\textbf{k},\alpha},A^{+}_{\textbf{k}',\beta}]=
\delta_{\alpha,\beta}\delta({\textbf{k}-\textbf{k}'})$.

The bound states of elementary particles (fermions) are associated with
bilocal quantum fields formed by the instantaneous potentials (see
\cite{MarkovYukawa,kad,LukierskiOziewicz})
 \bea\label{set-1} &&{\cal
M}(x,y)={\cal M}(z|X)\\\nonumber &=&\sum\limits_H\int\frac{d^3{\cal
P}}{(2\pi)^{3}\sqrt{2\omega_H}}\int\frac{d^4qe^{iq\cdot z}}{(2\pi)^4} \{
e^{i{\cal P}\cdot{X}} \Gamma_H(q^{\bot}|{\cal P})a^+_H(\bm{{\cal
P},q^{\bot}})+e^{-i{\cal P}\cdot{X}} \bar{\Gamma}_H(q^{\bot}|{\cal
P})a^-_H(\bm{{\cal P},q^{\bot}})\},
 \eea
where ${\cal P}\cdot X=\omega_H X_0-\bm{{\cal P}} \mathbf{X}$, ${\cal
P}_\mu=(\omega_H,\bm{{\cal P}})$ is the total momentum components on the mass
shell (that is, $\omega_H=\sqrt{M_H^2+\bm{{\cal P}}^2}$), and \bea
\label{set-1a} X=\frac{x+y}{2},\ \ z=x-y. \eea are the the total coordinate
and the relative one, respectively. The functions $\Gamma $ belongs to the
complete set of orthonormalized solutions  of the BS equation \cite{a15} in a
specific gauge theory, $a^{\pm}_H(\bm{{\cal P},q^{\bot}})$ are coefficients
treated in quantum theory as the creation (+) and annihilation operators (see
Appendix B).

The irreducibility constraint, called Markov-Yukawa constraint, is imposed on
the class of instantaneous bound states
 \bea \label{set-2}
 z^\mu \hat {\cal P}_\mu{\cal M}(z|X)\equiv i z^\mu \frac{d}{d
X^\mu}{\cal M}(z|X)=0.
  \eea

In Ref.~\cite{logunov} the in- and out- asymptotical states are the ``rays''
defined as a product of these irreducible representations of the Poincar\'e
group \be \label{3-2} \langle {\rm out}|=\langle \prod_{J}{{\cal
P}_J,s_J}~\big|, ~~~|{\rm in}\rangle =\big|\prod_{J }{{\cal P}_J,s_J}\rangle.
\ee This means that all particles (elementary and composite)  are  far enough
from each other to neglect their interactions in the in-, out- states. All
their asymptotical states  $\langle {\rm out}|$ and $|{\rm in}\rangle $
including the bound states are considered as the irreducible representations
of the Poincar\'e group.

These irreducible representations form a complete set of states, and the
reference frames are distinguished by the eigenvalues of the appropriate time
operator $\hat \ell_\mu=\dfrac{\hat{\cal P}_\mu}{M_J}$
 \bea \label{3-2a}
 \hat \ell_\mu|{{\cal P},s}\rangle =\frac{{\cal P}_{J\mu}}{M_{J}}|{{\cal P}_J,s}\rangle ,
 \eea
  where the Bogoliubov-Logunov-Todorov  rays (\ref{3-2a}) can include bound
states.

\section{Symmetry of S-Matrix
}\label{sect_P}

The S-matrix elements are defined as the evolution operator expectation values
between in- and out- states
 \bea \label{4-2}\underbrace{{{\cal M}_{\rm in,out}}}_{P-inv,G-inv}=
 \underbrace{\langle {\rm out}|}_{P-variant}~
 \underbrace{\hat S[\hat \ell]}_{P-variant,G-inv}~\underbrace{|{\rm in}\rangle }_{P-variant},
\eea where the abbreviation ``G-inv'', or ``gauge-invariant'', assumes the
invariance of S-matrix with respect to the gauge transformations.

The Dirac approach to gauge-invariant S-matrix was formulated at the rest
frame $\ell_\mu^0=(1,0,0,0)$ \cite{Dirac,hp,Polubarinov}. Then the problem
arises how to construct a gauge-invariant S-matrix in an arbitrary frame of
reference. It was  Heisenberg and Pauli's question   to von Neumann: ``How to
generalize the Dirac Hamiltonian approach to QED of 1927 \cite{Dirac} to any
frame?'' \cite{Pervu2,hp,Polubarinov,Pervu}. The  von Neumann reply was to go
back to the initial Lorentz-invariant formulation  and to choose the comoving
frame \be\label{vN-1} \ell_\mu^0=(1,0,0,0) \to \ell_\mu^{\rm
comoving}=\ell_\mu,~~~~~\ell_\mu \ell^\mu=\ell\cdot \ell=1 \ee and to repeat
the gauge-invariant Dirac scheme in this frame.

 Dirac Hamiltonian
approach to QED of 1927 was based on the constraint-shell action \cite{Dirac}
 \bea \label{6-2} W^{\rm Dirac}_{\rm QED}=W_{\rm QED} \Big|_{\frac{\delta
 W_{\rm QED}}{\delta A^{\ell}_0}=0},
  \eea
   where the component $A^{\ell}_0$ is
defined by the scalar product $A^{\ell}_0=A\cdot {\ell}$ of vector field
$A_\mu$ and the unit time-like vector ${\ell}_\mu$.

The gauge was established by Dirac as the first integral of the Gauss
constraint
 \bea \label{7-2} \int^t dt{\frac{\delta W_{\rm QED}}{\delta
A^{\ell}_0}=0},~~~~~~~~ t=(x\cdot \ell).
 \eea
In this case, the S-matrix
elements (\ref{4-2}) are relativistic invariant and independent of the frame
reference provided the condition (\ref{3-2a}) is fulfilled
\cite{LukierskiOziewicz,Kalinovsky}.

Therefore, such relativistic  bound states can be successfully included in the
relativistic covariant unitary perturbation theory \cite{Kalinovsky}. They
satisfy the Markov-Yukawa constraint (\ref{set-2}).

This framework yields the observed spectrum of bound states in QED
\cite{Salpeter}, which corresponds to the instantaneous potential interaction
and paves a way for constructing a bound state generating functional. The
functional construction is based on the Poincar\'e group representations
(\ref{fp-2d}) with $\ell^{0}$ being the eigenvalue of the total momentum
operator of instantaneous bound states.

 %instead of  of the rest frame
\section{QED}\label{sect_B}

\subsection{Split of potential part from radiation one}
 Let us  formulate the statement of the bound state problem in the terms of the
 gauge-invariant variables using QED.
 It is given by the action \cite{Polubarinov}
 \begin{eqnarray}\label{w}
 W[A,\psi,\bar \psi] &=& \int {d}^4 x \,\,\, \Bigl[ -\frac{1}{4} ( F_{\mu\nu}
 )^{2} + \bar {\psi} ( i \,\, \rlap/\nabla (A) - m^{0} ) \psi
 \Bigr]  \,\,\, ,
 \end{eqnarray}
 %Form(4.2)
 \begin{eqnarray}\label{vw}
 \nabla_{\mu}(A) &=& \partial_{\mu} -i e {A}_{\mu}, \,\,\,\,\,
 \rlap/\nabla = \nabla_\mu \cdot \gamma^\mu \,\,\, ,
 \nonumber \\
 \,\,\,\,\,F_{\mu\nu} &=&
 \partial_{\mu}A_{\nu} - \partial_{\nu}A_{\mu}~.
 \end{eqnarray}

 Dirac defined these gauge-invariant variables
 by the transformations
 \begin{eqnarray}\label{vak}
 \sum_{a=1,2} e_k^aA_a^D=A_{k}^{D}[A]& =&
 v[A] \Big( A_{k} + i \frac{1}{e} \partial_{k} \Big)
 ( v[A] )^{-1},\\
 \psi^{D}[A,\psi]& =& v[A] \psi \,,\,\,
 \end{eqnarray}
 where the gauge factor  is given by
 %2.5
 \bea\label{va}
 v[A] &=&  \exp \bigl\{ ie\int_{}^{t} dt' a_0 (t')\bigr\}~\\
 \label{var2s}
 a_0[A]&=&\frac{1}{\Delta}
 \partial_i  \partial_0 A_i(t,x)~.
 \eea
 Here the inverse Laplace operator acts on arbitrary function $f(t,\mathbf{x})$ as
\bea
\label{red-3}
\dfrac{1}{\Delta}f(t,\mathbf{x})&\stackrel{\rm
def}{=}&-\frac{1}{4\pi}\int \textrm{d}^3 y
\frac{f(t,\mathbf{y})}{|\;\mathbf{x}-\mathbf{y}|}
\eea
with the kernel being the Coulomb potential.

 Using the gauge transformations~%(\ref{gauge})
 \be\label{vgauge}
 a^{\Lambda}_0=a_{0}+\partial_0 \Lambda~\Rightarrow~
 v[A^{\Lambda}]= \exp[ie\Lambda(t_0,\mathbf{x})]v[A]\exp[-ie\Lambda(t,\mathbf{x})]~,
 \ee
 we can find that  initial data of
 the gauge-invariant Dirac variables~(\ref{vak}) are
 degenerated with respect to the stationary gauge transformations
  \be\label{gauge1}
 A_i^D[A^\Lambda]\;=\;A_i^D[A]+\partial_i \Lambda(t_0,\mathbf{x}),~~~~~
 \psi^{D}[A^\Lambda,\psi^\Lambda]
 =\exp[ie\Lambda(t_0,\mathbf{x})]\psi^{D}[A,\psi]~.
 \ee
 The Dirac variables (\ref{vak}) as the functionals of the initial fields
 satisfy the Gauss law constraint
 \be\label{gc2}
 \partial_0 \left(\partial_i A^D_i(t,\mathbf{x}) \right) \equiv 0~.
 \ee
 Thus, explicit resolving the Gauss law allows us
 to remove two degrees of freedom and to reduce
 the gauge group into the subgroup of the stationary gauge
 transformations~(\ref{gauge1}).

 We can fix a stationary phase $\Lambda(t_0,\mathbf{x})=\Phi_0 (\mathbf{x})$
 by  an additional constraint in the form of the
 time integral of the Gauss law constraint~(\ref{gc2})
 with zero initial data
 \be \label{ngauge}
 \partial_i A_i^D=0 \to \Delta\Phi_0 (\mathbf{x})=0.
 \ee
 Dirac constructed  the {\it unconstrained system}, equivalent to the initial theory~(\ref{w})
 \bea \label{red}
 &&W^*= W|_{\delta W/{\delta A_0}=0}[A_a^D=A_a^*,\psi^D=\psi^*]\\\nonumber
 &&=\int d^4x
 \left[\frac{(\dot A_i^*)^2-B_i^2}{2} %\sum_{a=1,2}(\partial_\mu% A_a^*\partial^0 A_a^*)
 + \frac{1}{2}j_{0}^{*} \frac{1}{\Delta}j_{0}^{*} -
 j_{i}^{*} A_{i}^{*} + {\bar {\psi}}^{*} (i{\hat \partial} -
 m){\psi}^{*}\right]\;,
  \eea
where
\bea\label{red-1} \dot A_{i}^{*}&=&\sum\limits_{a=1,2}\partial_0A^*_{\mathrm{a}}e_{i}^{a},\\\label{red-2}
B_i&=&\varepsilon_{ijk}\partial_j A_{k}^{*}
\eea
are the electric and magnetic fields, respectively.

In three-dimensional QED, there is a subtle difference between the model~(\ref{red}) and the initial gauge theory~(\ref{w}). This is the origin of the current conservation law. In the initial constrained system~(\ref{w}), the current conservation law  $\partial_{0}j_{0}=\partial_{i}j_{i}$ follows from the equations for the gauge fields, whereas a similar law $\partial_{0}j^*_{0}=\partial_{i}j^*_{i}$ in the {\it equivalent unconstrained  system}~(\ref{red}) follows only from the classical equations for the fermion fields. This difference becomes essential  in quantum theory. In the  second case, we cannot use the current conservation law if the quantum fermions are off mass-shell, in particular, in a bound state. What do we observe in an atom? The bare fermions, or {\it dressed} ones~(\ref{vak})? Dirac supposed~\cite{Dirac} that we can observe only {\it gauge invariant} quantities of the type of the {\it dressed} fields.

\subsection{Bilocal fields in the Ladder Approximation  %as unitary representations of the Poincar\'e group
}

The  constraint-shell QED allows us to construct the relativistic covariant perturbation theory with respect to radiation corrections ~\cite{love}. Recall that our solution of the problem of relativistic invariance of the nonlocal objects is the choice of the time axis as a vector operator with eigenvalues proportional to the total momenta of  bound states \cite{yaf,a20}. In this case, the relativistic covariant unitary S-matrix can be defined as the Feynman path integral
\be \label{fi2}
  Z_{\hat \eta}^{*}[ s, {\bar {s}}^*, J^* ]\;=\langle *|\int
  D\psi^*D{\bar \psi}^*
  e^{iW_{\hat \eta}^{*}[\psi^*, {\bar \psi}^*] + i S^* }|*\rangle ~,
  \ee
where
\begin{eqnarray} \label{act2}
  W_{\hat \eta} [ \psi , \bar{\psi} ] &=&
  \int d^{4}x [ \bar{\psi}(x) ( i \rlap/{\partial}- ie\rlap/A^* - m^{0} ) \psi(x) +   \nonumber  \\
  &+& { 1 \over 2 } \int d^{4}y ( \psi(y) \bar{\psi}(x) ) {\cal
  K}^{(\ell)} ( z^{\bot} \mid X ) ( \psi(x) \bar{\psi}(y) ) ] ,
\end{eqnarray}
and the symbol
  \be
   \langle *|\ldots|*\rangle =\;\int \prod_{j} DA^*_je^{iW_0^{*}[A^*]}\ldots
\ee
stands for the averaging over transverse photons.
Here by definition $\rlap/\partial = \partial^{\mu} \gamma_{\mu}$, and ${\cal
K}^{(\ell)}$ is the kernel
\begin{align} \label{3-7}
  &{\cal K}^{(\ell)}( z^{\perp} \mid X ) = \rlap/\ell V(z^{\perp})
   \delta(z \cdot \ell ) \rlap/\ell,\\
  &\rlap/\ell = \ell^{\mu}
  \gamma_{\mu}=\gamma \cdot \ell,~~~~ z_{\mu}^{\perp} = z_{\mu} - \ell_{\mu}
  ( z \cdot \ell ), \nonumber
\end{align}
where $z$ and $X$ are the relative and total coordinates (\ref{set-1a}). The potential $ V( z^{\perp} ) $ depends only on the transverse  component of the relative coordinate with respect to the time axis $\ell$. The requirement for the choice of the time axis (\ref{3-2a}) in
bilocal dynamics is equivalent to Markov - Yukawa  condition (\ref{set-2}).

Apparently, the most straightforward way for constructing a theory of bound states is the redefinition of action (\ref{act2}) in terms of the bilocal fields by means of the Legendre transformation \cite{pre-1a}
\begin{eqnarray}  \label{3-8}
  {1 \over 2} &\int &d^{4}x d^{4}y  ( \psi(y) \bar{\psi}(x) ) {\cal
  K}(x,y) ( \psi(x) \bar{\psi}(y) )  = \nonumber \\  = -{1 \over 2}
  &\int &d^{4}x d^{4}y  {\cal M}(x,y) {\cal K}^{-1}(x,y) {\cal
  M}(x,y) + \\  +  &\int &d^{4}x d^{4}y ( \psi(x) \bar{\psi}(y) ),
  {\cal M}(x,y) \nonumber
\end{eqnarray}
where $ {\cal K}^{-1} $ is the inverse kernel $ {\cal K}$ given by Eq.~(\ref{3-7}). Following Ref.~\cite{pre-1}, we introduce the short-hand notation
\begin{eqnarray} \label{short}
  \int d^{4}x d^{4}y \psi (y) \bar{\psi}(x) ( i
  \rlap/\partial -ie\rlap/A^*- m^{0} ) \delta^{(4)} (x-y)
  &=&  ( \psi \bar{\psi} , - G_{A}^{-1} ) \, \, ,  \\
  \int d^{4}x d^{4}y ( \psi(x) \bar{\psi}(y) ) {\cal M}(x,y) &=& (
  \psi \bar{\psi}, {\cal M} ) .
\end{eqnarray}

After quantization over $N_{c}$ fermion fields (here $N_c$ is the number of colors equal to 3), the functional~(\ref{fi2}) takes the form
  \be
  \label{fi3}
  Z_{\hat \eta}^{*}[ s, {\bar {s}}^*, J^* ]\;=\langle *|\int \prod D{\cal M}
  e^{iW_{eff}[{\cal M}] + i S_{eff}[{\cal M}]}|*\rangle,
  \ee
  where
  \begin{eqnarray} \label{3-9}
  W_{eff}[{\cal M}] = %\psi \bar{\psi},
  {\rm tr}\left[\log ( - G_{A}^{-1} + {\cal M})
  \right] - { 1 \over 2} ( {\cal M}, {\cal K}^{-1} {\cal M} )
  \end{eqnarray}
is the effective action, and
\begin{eqnarray} \label{3-10}
S_{eff}[{\cal M}] = ( s^* \bar{s^*}, (  G_{A}^{-1} - {\cal
M})^{-1} )
 \end{eqnarray}
is the source term. The effective action can be decomposed as
 \begin{eqnarray} \label{3-11}
W_{eff}[{\cal M}] = - {1 \over 2}  ( {\cal M}, {\cal K}^{-1}
{\cal M} ) + i \sum_{n=1}^{\infty} {1 \over n} \Phi^{n} .
\end{eqnarray}
Here $ \Phi \equiv G_{A} {\cal M} , \Phi^{2} , \Phi^{3} $, etc.
mean the following expressions
\begin{eqnarray}   \label{3-12}
\Phi (x,y) \equiv G_{A} {\cal M} = \int d^{4}z G_{A} (x,z) {\cal M}(z,y), \nonumber \\
\Phi^{2} = \int d^{4} x d^{4}y \Phi(x,y) \Phi(y,x) , \\
\Phi^{3} = \int d^{4} x d^{4}y d^{4}z \Phi(x,y) \Phi(y,z)
\Phi(z,x), \,\,  \text{etc.} \nonumber
\end{eqnarray}

As a result of such quantization, only the contributions with inner fermionic
lines (but not the scattering and dissociation channel contributions) are
included in the effective action, since we are interested only in the bound
states constructed as unitary representations of the Poincar\'e group.

\subsection{The anomalous creation of Para-Positronium in QED}

The effective bound state functional in the presence of radiation fields
contains a triangle anomaly decay of positronium $\eta_P$ with an additional time derivative of
these fields \bea \label{s-11}\nonumber
  W_{eff}&=&{W(A^*)}+{W(\eta_P)}%+W_{\rm anomaly}
  \\\nonumber
{W(A^*)}&=&\int d^4x  [{C_P\eta_P \dot A_i B_i}+
{\frac{\dot A^2_i +B^2_i}{2}}]\\\nonumber
  W_{\eta}&=&\int d^4x \left\{\frac 1 2 \left[{\dot\eta_P}^2\!-\!M_L^2{\eta_P}^2\!-
  \left(\!\partial_i\eta_P\!\right)^2
  \right]\!\right\},
  \eea
  where
  $B_i$ are the magnetic field component (\ref{red-2}), and the parameter of the effective action is given by
  \bea \label{t-1}%\nonumber
  C_P&=&\frac{2\alpha}{m_e}\left(\frac{\underline{\psi}_{\mathrm{Sch}}(0)}{m_e^{3/2}}\right)=
   \frac{\sqrt{\pi}\alpha^{5/2}}{m_e}=\dfrac{\alpha\sqrt{2}}{F_P\sqrt{\pi}},
  \eea
where $\alpha=1/137$ is the QED coupling constant, and
\bea \label{t-2a}%\nonumber
  F_P&=&\frac{\underline{\psi}_{\mathrm{Sch}}(0)}{m_e^{3/2}}
  \frac{m_e}{\sqrt{2\pi}}
  = \frac{m_e\sqrt{2}}
  {\alpha^{3/2}\pi}
  \eea
is the positronium analogy of the pion weak-coupling constant $F_\pi$
discussed in the next Section.

\begin{figure}[tbh]
\begin{center}
\includegraphics{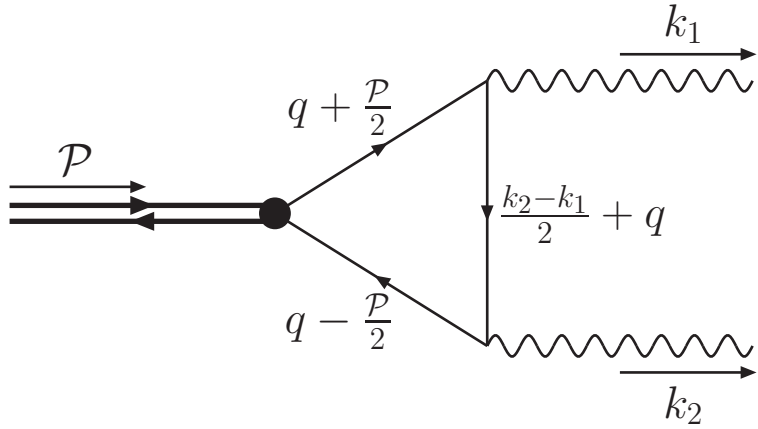}
\end{center}
\caption{\label{fig:1} The standard triangle diagram, used for calculating $C_P$, the parameter of the effective action.}
  % 4) Формула (53),
%проверьте индексы, в левой части буква b стоит 3 раза, d - один.}
\end{figure}

 The product   $C_P\dot A_i B_i$ is obtained
  from the triangle diagram shown in Fig.~\ref{fig:1}.

The Hamiltonian of this system is the sum of the Hamiltonians of the free electro-magnetic fields and the
    positronium ones $\eta_P(x)$ and the interaction
 \bea \label{s-12}\nonumber
  W_{eff}&=&\int dt d^3x \left\{
  E_i \dot A_i + P_\eta \dot \eta_P-{\cal H}\right\},\\\nonumber
  {\cal H}&=&
  {\cal H}_\eta+
  {\cal H}_A+{\cal H}_{\rm int},\\\label{s-13}\nonumber
  {\cal H}_\eta&=&\frac 1 2 \left({\dot\eta_P}^2-M_P^2{\eta_P}^2-(\partial_i{\eta_P})^2\right),\\
 \label{s-14}%\nonumber
 {\cal H}_{\rm int }
&=&C^2_P\eta_P  E_i B_i+{\frac{C^2_P\eta^2_P}{2}B_i^2}~.
  \eea
\begin{figure}[tbh]
\begin{center}
\includegraphics{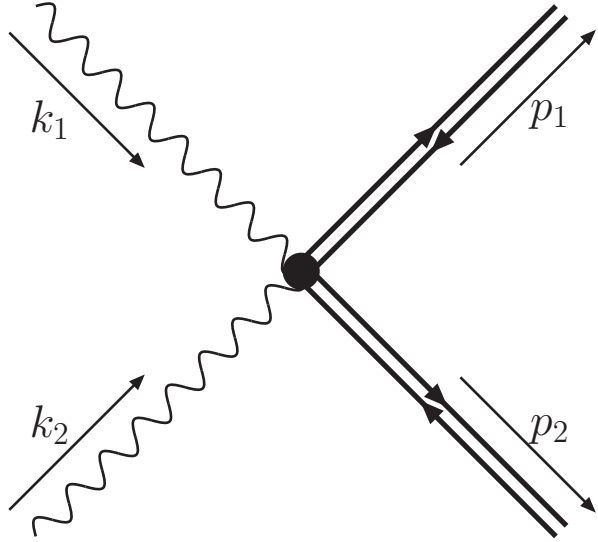}
\end{center}
\caption{\label{fig:2} The new diagram for the anomalous processes of creation
of the positronium pairs in the external magnetic field that follows from the Hamiltonian anomaly, given by
${\cal H}_{\rm int }$ in Eqs.~(\ref{s-12}).}
\end{figure}
The anomalous processes of creation
of the positronium pairs in the external magnetic field
 at the photon energy value $E_\gamma\simeq m_e $ (see Fig.~\ref{fig:2}) are described by
 the differential cross section
\bea\label{cp} \frac{d\sigma}{d\Omega} =\frac{\alpha^{4}E_\gamma^2}{\pi 128
F_P^4} \sqrt{1-{\left(\frac{2 m_e}{E_\gamma}\right)}^2} =\frac{\pi
\alpha^{10}E_\gamma^2}{512 m_e^4} \sqrt{1-{\left(\frac{2
m_e}{E_\gamma}\right)}^2}. \eea

\subsection{The Schwinger  QED${}_{1+1}$}

The Schwinger two dimensional QED${}_{1+1}$  was considered in the framework of the  Dirac approach to gauge theories distinguished by the constraint-shell action \cite{pvn7}.

This constraint-shell action has an additional time derivative term of the gauge field that goes from the fermion propagator in the axial anomaly. This anomalous time derivative term  changes the initial Hamiltonian structure of the gauge field action
 \bea \label{sc-11}
  W_{\rm Scwinger}&=&\int dt dx \left\{\frac 1 2 {\dot\eta_S}^2
   +
  C_S\eta_S \dot A+\frac{\dot A^2}{2}
   \right\}\\
   \label{sc-12}  &=&\int dt dx \left[P_S \dot \eta_S+E\dot A-\frac{E^2}{2}
   +
  C_S\eta_S E-C^2_S\frac{\eta^2_S}{2}\right],\\\label{sc-13}
  C_S&=&\frac{e}{2\pi}.
  \eea
Finally,
an additional Abelian anomaly given by the last term in Eq.~(\ref{sc-12}) enables us to determine the mass of the pseudoscalar bound
state~\cite{pvn7}.
 In $QED_{(1+1)}$, it is the well-known mass of the Schwinger bound state
 $$
  \triangle{M}^2 %=\frac{C_{M}^2}{I_{\rm QED}V}
  =\frac{e^2}{\pi}.
 $$
The Schwinger model justifies including of the similar additional terms
in the 4-dimensional QED.

%\appendix
\section{Non-Abelian Dirac Hamiltonian Dynamics in an Arbitrary Frame of Reference}  \label{sect_A}

In order to demonstrate the Lorentz-invariant version of the
  Dirac method \cite{Dirac} given by Eq.(\ref{6-2})
 in a non-Abelian theory, we
 consider
 the simplest example of the Lorentz-invariant  formulation of the naive
 path integral without any ghost fields and FP-determinant
 %The von Neumann idea \cite{hp} to give the Lorentz-invariant form of
% the Dirac Hamiltonian dynamics in an arbitrary frame of reference \cite{Dirac}
% %given by Eq.(\ref{6-2})
%   can be applied in a non-Abelian theory.
% Recall that
% the simplest example of the Lorentz  formulation that uses the naive
% path integral without any ghost fields and FP-determinant is described by the
% generating functional
  \bea\label{fp-1c}
 Z[J,\eta,\overline{\eta}]&=&\int\left[\prod\limits_{\mu,a}^{}d A^a_\mu\right]
 d\psi d\overline{\psi} e^{iW[A,\psi,
 \overline{\psi}]+iS[J,\eta,\overline{\eta}]}.
 \eea
 We use standard the QCD action $W[A,\psi, \overline{\psi}]$ and the source
terms \bea \label{fp-2} W&=&\int d^4x
\left[-\frac{1}{4}F^a_{\mu\nu}{F^a}^{\mu\nu}
-\overline{\psi}(i\gamma^\mu(\partial_\mu +\hat A_\mu) -m)\psi\right],\\
\label{fp-4} F^a_{0k}&=&\partial_0 A^a_k -\partial_0
A^a_k\partial+gf^{abc}A^b_0 A^c_k\equiv
\dot A^a_k-\nabla^{ab}_k A^b_0,\\
\label{fp-42} S&=&\int d^4x \left[A_\mu
J^\mu+\overline{\eta}\psi+\overline{\psi}\eta\right], ~~\hat
A_\mu=ig\frac{\lambda^aA^a_\mu}{2}. \eea

  %where the action $W[A,\psi,
%\overline{\psi}]$ is given by Eq.(\ref{fp-2}) %We use standard the QCD action
%and the source terms
%\bea \label{fp-2}
%W&=&\int d^4x \left[-\frac{1}{4}F^a_{\mu\nu}{F^a}^{\mu\nu}
%-\overline{\psi}(i\gamma^\mu(\partial_\mu +\hat A_\mu) -m)\psi\right],\\
%\label{fp-4}
%F^a_{0k}&=&\partial_0 A^a_k -\partial_0 A^a_k\partial+gf^{abc}A^b_0 A^c_k\equiv
%\dot A^a_k-\nabla^{ab}_k A^b_0,\\
%\label{fp-42}
%S&=&\int d^4x \left[A_\mu J^\mu+\overline{\eta}\psi+\overline{\psi}\eta\right],
%~~\hat A_\mu=ig\frac{\lambda^aA^a_\mu}{2}.
%\eea

There are a lot of drawbacks of this path integral from the point of view of
the Faddeev-Popov functional \cite{fp1}. They are the following:
\begin{enumerate}
\item The  time component $A_0$ has indefinite metric.
\item The integral (\ref{fp-1c}) contained the infinite gauge factor.% due to pure gauge degrees of freedom.
\item The bound state spectrum contains tachyons.
\item The analytical properties of  field propagators are gauge dependent.
\item Operator foundation is absent \cite{sch}.
\item Low-energy region does not separate from the high-energy one.
\end{enumerate}

All these defects can be
removed by the integration over the indefinite metric  time component $A_\mu \ell^\mu\equiv A\cdot \ell$,
 where $\ell^\mu$ is an arbitrary
unit time-like vector: $\ell^2=1$. If $\ell^{0}=(1,0,0,0)$ then $A_\mu
\ell^\mu= A_0$. In this case
 \bea
 \label{fp-2d}
 Z[{{\ell^0}}]&=&\!\int\!\!  \left[\prod\limits_{x,j,a}^{}dA^{a*}_j(x)\!\right]\!e^{iW^*_{\rm YM}}
\! \delta\left(L^a\right) \left[\det{(\nabla_j(A^*))^2}\right]^{-1/2}Z_{\psi},\\
\label{fp-2id}
L^a&=&\int\limits_{}^{t} d\overline{t}\nabla^{ab}_i(A^*)\dot A^{*b}_i=0,\\
\label{fp-3id}
W^*_{\rm YM}&=&\int d^4x \frac{(\dot {A_j^a}^*)^2-(B^a_j)^2}{2},\\
\label{fp-3d} Z_{\psi}&=& \int d\psi
d\overline{\psi}e^{-\frac{i}{2}\left(\psi\overline{\psi},{\cal
K}\psi\overline{\psi}\right)-\left(\psi \overline{\psi},G^{-1}_{A^*}\right)
+i S[J^*,\eta^*]}\\
%\label{eff-1}!!!!!!!!!!!!!!!!!!!!!!!!!!!!!!!!!!!!!!!!!!!!!!!!!!!!!!!!!!!!!!!!!!!!!!!!!!!!!!!!!!!!!!!!!!!!!!!!!!!!!!!!!!!!!!!!!!!!!!!!!!!!!!!!!!!!!!!!!!!!!!!!!!!!!!!!!!!!!!!!!!!!!!!!!
%&=&\int\left[\prod\limits_{x,y,a,b}^{} d {\cal M}^{ab}(x,y)\right]\exp\{iW_{\rm eff}[{\cal M},A^*]+i(\eta\overline{\eta},G_{{\cal M}})\},\\
\label{fp-3iG}
\left(\psi \overline{\psi},G^{-1}_{A^*}\right)&=&\int d^4x\overline{\psi}\left[i\gamma_0\partial_0\!-\!\gamma_j(\partial_j+\hat A^*_j)\!\!-m\!\right]{\psi},\\
\label{fp-4id} \left(\psi\overline{\psi},{\cal K}\psi\overline{\psi}\right)&=&
\int d^4x d^4y
j^a_0(x)\left[\frac{1}{(\nabla_j(A^*))^2}\delta^4(x-y)\right]^{ab} j^b_0(y).
 \eea
 The infinite factor is removed by the gauge fixing (\ref{fp-2id}) treated
as an antiderivative function of the Gauss constraint. $A^{*a}_i$ denotes
fields $A^{a}_i$ under gauge fixing condition (\ref{fp-2id}). It becomes
homogeneous $\nabla^{ab}_i(A^{*})\dot A^{*b}_i=0$ because $A^*_0$ is
determined by the interactions of currents (\ref{fp-4id}).
 It is just the non-Abelian
generalization \cite{Pervu2,inp,Pervu,npbs} of the Dirac approach to QED \cite{Dirac}.
In the case of QCD there is a possibility to include the nonzero condensate
of transverse gluons $\langle A^{*a}_jA^{*b}_i\rangle=2C_{\rm gluon}\delta_{ij}\delta^{ab}$.

The Lorentz-invariant bound state matrix elements can be obtained, if we choose
the time-axis $\ell$
of Dirac Hamiltonian dynamics  as the operator
acting in the complete set of bound states (\ref{3-2a})
and given by  Eqs.(\ref{set-1a}) and (\ref{set-2}).
 %{This scheme enables us to
%restore the Lorentz-invariance, if  the time-axis as the operator $\hat\ell$
 %proper frame of reference of each bound states using the
 This means the von Neumann substitution (\ref{vN-1}) given in \cite{hp}
\begin{equation}\label{3-2c}
Z_{}[\ell^{0}\,]\to Z[\,\ell\,]\to Z[\,{\hat \ell}\,]
\end{equation}
instead of the Lorentz-gauge formulation \cite{fp1}. %\cite{faddeev69}
%acting in the complete set of bound states.}

\section{Axial Anomalies in the NJL model inspired by QCD}\label{sect_H}

\subsection{Formulation of the NJL model inspired by QCD}

Instantaneous QCD interactions are described by the non-Abelian generalization
of the Dirac gauge in QED
%The relativistic invariant bilocal effective action obtained in \cite{yaf},
%takes for the quark sector in the color singlet channel the form
\bea\label{1-1} S_{\rm inst}  &=& \int d^{4} x  \bar {q}(x) \; ( i
\rlap/\partial - {\hat {m}}^{0} ) {q}(x)-{ 1 \over {2}}
\int d^4x d^4y j^a_0(x)\left[\frac{1}{(\nabla_j(A^*))^2}\delta^4(x-y)\right]^{ab} j^b_0(y)
%\\\nonumber
% &-&   { 1 \over {2}} %N_{c}
%\int d^{4} y \;\;
%q_{2'}(y) \bar {q}_1(x) \;\;
%{\cal K}_{(1,1'|2,2')}^{\ell} (x - y) \;\;
%q_{1'}(x) \bar {q}_{2}(y) \;\; \} \;\; ,
\eea
where $j^a_0(x)=\bar q(x)\dfrac{\lambda^a}{2}\gamma_0 q(x)$ is the 4-th component of the quark current,
with the Gell-Mann color matrices $\lambda^a$
(see the notations in Appendix A).
%For  simplicity indices $(1,1'|2,2')$ denote in (1) all spinor, color and flavor
%ones.
The symbol
$
{\hat {m}}^{0} = \text{diag}(m_{u}^{0}, m_{d}^{0},m_{s}^{0})
$
denotes the bare quark mass matrix.

%where  the parameter $M_g$ is defined in the perturbation theory by the {fp-4id}
The normal ordering of the transverse gluons in the nonlinear action (\ref{fp-4id})
$\nabla^{db}A^b_0\nabla^{dc}A^c_0$
leads to
the condensate of gluons
\bea\label{c-1g} g^2f^{ba_1d}f^{da_2c}\langle
 A^{a_1*}_iA^{a_2*}_j\rangle &=&2g^2[N_c^2-1] \delta^{bc} \delta_{ij}C_{\rm gluon}=M_g^2
 \delta^{bc}\delta_{ij},\eea
where
 \bea\label{c-1gc}
\langle A^{*a}_jA^{*b}_i\rangle&=&2C_{\rm gluon}\delta_{ij}\delta^{ab}.
 \eea
This condensate yields the squired effective gluon mass in  the squared covariant derivative
$\nabla^{db}A^b_0\nabla^{dc}A^c_0=:\nabla^{db}A^b_0\nabla^{dc}A^c_0: + M_g^2
A^d_0A^d_0$ of constraint-shell action (\ref{fp-4id}) given in Appendix A. The
constant
$$C_{\rm gluon}=\int\dfrac{ d^3k}{(2\pi)^32\sqrt{{\bf k}^2}}$$
is finite after substraction of the infinite volume contribution,
and its value is determined by the hadron size like
the Casimir vacuum energy \cite{sh-79}.
Finally, in the lowest order of perturbation theory,
this gluon condensation  yields  the effective Yukawa potential in the colorless meson sector
\be\label{Y-1}
\underline{V}({\mathbf{k}})=\frac{4}{3}g^2\frac{1}{\mathbf{k}^2+M_g^2}
\ee
and the NJL type model with the effective gluon mass $M_g^2$. While deriving the last equation, we use the relation
\begin{eqnarray*}
\left[\sum\limits_{a=1}^{a=N_c^2-1}  \frac{\lambda^{a}_{1,1'}}{2}
\frac{\lambda^{a}_{2,2'}}{2}\right]_{\rm colorless}=\frac{4}{3}\delta_{1,2'}\delta_{2,1'}
% =\rlap/\ell \;\;V( z^{\perp} ) \;\; \delta (z \cdot \ell) \;\; \rlap/\ell \;\;
\end{eqnarray*}
in the colorless meson sector.

Below we consider the potential model (\ref{1-1}) in the form
\bea\label{1-1q} S_{\rm inst}  &=& \int d^{4} x  \bar {q}(x) \; ( i
\rlap/\partial - {\hat {m}}^{0} ) {q}(x)-{ 1 \over {2}}
\int d^4x d^4y j^a_{\ell}(x)V(x^{\bot}-y^{\bot}) \delta ((x-y)\cdot \ell) j^a_{\ell}(y)
%\\\nonumber
% &-&   { 1 \over {2}} %N_{c}
%\int d^{4} y \;\;
%q_{2'}(y) \bar {q}_1(x) \;\;
%{\cal K}_{(1,1'|2,2')}^{\ell} (x - y) \;\;
%q_{1'}(x) \bar {q}_{2}(y) \;\; \} \;\; ,
\eea
 with the  choice of the
time axis as the eigenvalues of the bound state total momentum, in the
framework of the ladder approximation given in Appendix A.
% of the operator

%All related formulas are  %{\color{red} We discuss only
%the effect of the}

 \subsection{Schwinger-Dyson equation: the fermion spectrum}

 The equation  of stationarity (\ref{8})   can be rewritten from the
  SD equation
 \begin{eqnarray}\label{sd}
 \Sigma(x-y) = m^{0} \delta^{(4)} (x-y) + i {\cal K}(x,y)
 G_{\Sigma}(x-y)~.
 \end{eqnarray}
 It describes the spectrum of Dirac
  particles in bound states.
 In the momentum space  with
 $
 \underline{\Sigma}(k) = \int d^{4}x \Sigma(x) e^{i k\cdot x}
 $
  for the Coulomb type kernel, we obtain the following equation for
 the mass operator (
 $ \underline {\Sigma} $ )
 \begin{eqnarray} \label{sdm}
 \underline{\Sigma}(k) = m^{0} + \frac{i}{2} \int { d^{4}q \over { (2\pi)^{4}
 }} \underline {V} ( k^{\perp} - q^{\perp} ) \rlap/\ell
 \underline{G}_{\Sigma}(q) \rlap/\ell ,
 \end{eqnarray}
 where $ G_{\Sigma}(q) = ( \rlap/q - \underline{\Sigma}(q))^{-1} ,
 \underline {V} ( k^{\perp} ) $ is the Fourier representation of the
 potential, $ k^{\perp}_{\mu} = k_{\mu} - \ell_{\mu} ( k \cdot
 \ell) $
 is the relative
transverse momentum.
The quantity $ \underline{\Sigma} $
 depends only on the transverse momentum
 $\underline{\Sigma}(k) = \underline{\Sigma}(k^{\perp})$,
 because of the instantaneous form of the potential $\underline{V}(k^{\perp} )$. We can put
 \begin{eqnarray} \label{sd21}
 \underline{\Sigma}_{\mathrm{a}}(q) = %\rlap/q^{\perp} +
   E_{\mathrm{a}} ({\bf q})\cos 2 {\upsilon}_{\mathrm{a}}({\bf q})\equiv M_{{\mathrm{a}}}({\bf q}).
 %S^{-2}_{\mathrm{a}} (q^{\perp})
 \end{eqnarray}
 Here $M_{\mathrm{a}}({\bf q})$
 is the constituent quark mass
 and
 \begin{eqnarray} \label{sd3}
  \cos 2 {\upsilon}_{\mathrm{a}}({\bf q})&=&\frac{M_{\rm a}(\bf q)}{\sqrt{M^2_{\rm a}(\bf q)+{\bf q}^2}}
 \end{eqnarray}
  determines  the Foldy - Wouthuysen type
 matrix
 \bea\label{s-1}
 S_{\mathrm{a}}({\bf q})=\exp [({\bf q}\bm{\gamma}/q){\upsilon}_{\mathrm{a}}(q)]=\cos {\upsilon}_{\mathrm{a}}(q)+ ({\bf q}\bm{\gamma}/q)\sin {\upsilon}_{\mathrm{a}}(q)
 \eea
 with  the vector of Dirac matrices
 %, lying in the range $0\leqslant {\upsilon}_{\mathrm{a}}(q)\leqslant \pi/2$,
 $\bm{\gamma}=(\gamma_1,\gamma_2,\gamma_3)$ and some angle ${\upsilon}_{\mathrm{a}}(q)$.
 The fermion spectrum can be obtained by solving the SD
 equation (\ref{sdm}).
 It can integrated over the longitudinal momentum $q_{0}= (q \cdot
 \ell) $ in the reference frame $\ell^{0}=(1,0,0,0)$, where $q^{\perp}=(0,\mathbf{q})$.
    By using Eq.~(\ref{s-1}), the quark Green function can be presented in the form
 \begin{eqnarray} \label{sd4}
 \underline{G}_{\Sigma_{\mathrm{a}}} &=&
 [ q_{0} \rlap/\ell - E_{\mathrm{a}}(q^{\perp}) S_{\mathrm{a}}^{-2}(q^{\perp})]^{-1} = \nonumber \\
 & = & \left[ { \Lambda^{(\ell)}_{(+)\mathrm{a}} (q^{\perp}) \over { q_{0} -
  E_{\mathrm{a}}(q^{\perp}) +i \epsilon} } +
 { \Lambda^{(\ell)}_{(-)\mathrm{a}} (q^{\perp}) \over
 { q_{0} + E_{\mathrm{a}}(q^{\perp}) +i \epsilon} } \right] \rlap/\ell,
 \end{eqnarray}
 where
 \begin{eqnarray} \label{ope1}
  \Lambda^{(\ell)}_{(\pm)\mathrm{a}}(q^{\perp})= S_{\mathrm{a}}(q^{\perp})
 \Lambda^{(\ell)}_{(\pm)}(0) S_{\mathrm{a}}^{-1}(q^{\perp}) , \,\,\,
 \Lambda^{(\ell)}_{(\pm)}(0)= ( 1 \pm \rlap/\ell ) / 2
  \end{eqnarray}
 are the operators separating the states with positive ($ + E_{\mathrm{a}}
 $) and negative ($ - E_{\mathrm{a}} $) energies.
 As a result, we obtain the following equations for the one-particles energy $ E $
  and the angle $ {\upsilon} $ with the potential given by Eq.~(\ref{Y-1})
 \bea\label{sd1}
 E_{\mathrm{a}}(k^{\perp}) \cos 2 {\upsilon}_{\mathrm{a}}(k^{\perp}) &=& m^{0}_{\mathrm{a}} +
 {1\over2}\int { d^{3}q^{\perp}\over (2\pi)^{3} }
 {V}(k^{\perp}-q^{\perp}) \cos 2{\upsilon}_{\mathrm{a}}(q^{\perp}).
\eea
In the rest frame $\ell^{0}=(1,0,0,0)$ this equation takes the form
\bea\label{sd2}%\\\nonumber
M_{\mathrm{a}}(k)&=&
%E_{\mathrm{a}}(k) \cos 2 {\upsilon}(k) =
m^{0}_{\mathrm{a}} +
 {1\over2}\int { d^{3}q\over (2\pi)^{3} }
 \underline{V}(k-q) \cos 2{\upsilon}_{\mathrm{a}}(q).
  \eea
By using the integral over the solid angle
$$\int_{0}^{\pi} d\vartheta \sin\vartheta \dfrac{2\pi}{M_g^2+({\bf k}-{\bf q})^2}
=\int\limits_{-1}^{+1} d\xi\dfrac{2\pi}{M_g^2+{k}^2+{q}^2-2{k}{q}\xi}
=\frac{\pi}{kq}\ln \frac{M_g^2+(k+q)^2}{M_g^2+(k-q)^2}$$
and the definition of the QCD coupling constant $\alpha_s=4\pi g^2$, it can be rewritten as
 \bea\label{sd-2a}%\nonumber
M_{\mathrm{a}}(k)&=&m^{0}_{\mathrm{a}} + \frac{\alpha_s}{3\pi k}\int\limits_{0}^{\infty}
 dq \frac{q M_{\mathrm{a}}(q)}{\sqrt{M^2_{\mathrm{a}}(q)+q^2}}\ln
 \frac{M_g^2+(k+q)^2}{M_g^2+(k-q)^2}.
 \eea

The suggested scheme allows us to consider the SD equation (\ref{sd2}) in the limit when
 the bare current mass $m^{0}_{\mathrm{a}}$ equals to zero. Then the ultraviolet divergence is absent, and, hence, the renormalization procedure can be successfully avoided.

This kind of nonlinear integral equations was considered in the paper \cite{puz} numerically.
 The solutions show us that in the region $q\ll M_g$ the function $\cos2\upsilon_{\mathrm{a}}$ is
  almost constant $\cos2\upsilon_{\mathrm{a}} \simeq 1$, whereas in the region $q\gg M_g$ the
  function $\cos2\upsilon_{\mathrm{a}}(q)$ decays in accordance with the power law
  $(M_g/q)^{1+\beta}$. The parameter $\beta$ is a
  solution of the equation
\begin{equation}\label{beta1}
\alpha_{s}\frac{\cot(\beta\pi/2)}{1-\beta}=\frac{3}{2},
\end{equation}
lying in the range $0<\beta<2$. This equation has two roots for $0<\alpha_s<3/\pi$, the first, belonging to the interval $0<\beta_1<1$, and the second, related to the first one by $\beta_2=2-\beta_1$. At $\alpha_s=3/\pi$, the two solutions merges into $\beta=1$, and there is no root for larger values of the coupling constant. Equation (\ref{beta1}) can be obtained by means of linearization of Eq.~(\ref{sd2}) within  the range $q\gg M_g$, because in this range $M_{\mathrm{a}}(q)\ll q$. Thus, the solution for $\cos2\upsilon_{\mathrm{a}}(q)$ is a reminiscent of the step function. This result justifies the estimation of the quark and meson spectra in the separable approximation \cite{a20} in agreement with the experimental data. Currently, numerical solutions of the nonlinear  equation (\ref{sd-2a}) are under way, and the details of computations will be published elsewhere.

 \subsection{Spontaneous chiral symmetry breaking }

As discussed in the previous section, the SD equation (\ref{sd2}) can be rewritten in the form (\ref{sd-2a}). Once we know the solution of Eq.~(\ref{sd-2a}) for $M_{\mathrm{a}}(q)$, we can determine the Foldy-Wouthuysen angles $\upsilon_{\mathrm{a}},(\mathrm{a}=u,d)$ for u-,d- quarks with the help of relation (\ref{sd3}). Then the BS equations in the form (\ref{pse})
 \begin{align}
\label{bs-3}
 M_{\pi}L^{\pi}_{2}(\textbf{p})&=[E_u(p)+E_d(p)]L^{\pi}_{1}(\textbf{p})\!-\!
 \!\int\!\frac{d^3q}{(2\pi)^3}\underline{V}(\textbf{p}\!-\!\textbf{q})
 L^{\pi}_{1}(\textbf{q})[c^{-}(p)c^{-}(q)\!+\!\xi s^{-}(p)s^{-}(q)],\\
\label{bs-3-+}
 M_\pi L^{\pi}_{1}(\textbf{p})&=[E_u(p)+E_d(p)]L^{\pi}_{2}(\textbf{p})\!-\!\!
 \int\!\frac{d^3q}{(2\pi)^3}\underline{V}(\textbf{p}\!-\!
 \textbf{q})L^{\pi}_{2}(\textbf{q})[c^{+}(p)c^{+}(q)\!+\!\xi s^{+}(p)s^{+}(q)]
\end{align}
yield the pion mass $M_\pi$ and wave functions $L^{\pi}_{1}(\textbf{p})$ and $L^{\pi}_{2}(\textbf{p})$.
Here $m_u,m_d$ are the current quark masses, $E_{\mathrm{a}}=\sqrt{p^2+M^2_{\mathrm{a}}(p)}, (\mathrm{a}=u,d)$
are the u-,d- quark energies,
$\xi=(\textbf{p\,q})/pq$, and we use the notations
 \bea
E({\bf p})&=&E_{\mathrm{a}}({\bf p})+E_{\mathrm{b}}({\bf p})~,\label{E-1a}\\
{\tt c}^{\pm}(p)&=&\cos[\upsilon_{\mathrm{a}}(p) \pm \upsilon_{\mathrm{b}}(p)]~,\label{E-2a}\\
{\tt s}^{\pm}(p)&=&\sin[\upsilon_{\mathrm{a}}(p) \pm \upsilon_{\mathrm{b}}(p)]~.\label{E-3a}
\eea

The model is simplified in some limiting cases. Once the quark masses
 $m_u$ and $m_d$ are small and approximately equal, then Eqs.~(\ref{sd2}) and
 (\ref{bs-3}) take the form
\begin{align}\label{sd-3u}
 m_{\mathrm{a}}&=M_{\mathrm{a}}(p)-\frac{1}{2}\int\frac{d^3q}{(2\pi)^3}\underline{V}(\textbf{p}-\textbf{q})
 \cos 2{\upsilon}_u(q),\\
\label{bs-3u}
 \frac{M_{\pi}L^{\pi}_{2}(\textbf{p})}{2}&=E_u(p)L^{\pi}_{1}(\textbf{p})\!-\!
 \frac{1}{2}\!\int\!\frac{d^3q}{(2\pi)^3}\underline{V}(\textbf{p}\!-\!\textbf{q})
 L^{\pi}_{1}(\textbf{q}).
\end{align}
Solutions of equations of this type are  considered in the numerous papers~\cite{a4,a5,a6,a7,a10} (see also review~\cite{puz}) for different potentials. One of the main results of these papers was the pure quantum effect of spontaneous chiral symmetry breaking. In this case, the instantaneous interaction leads to rearrangement of the perturbation series and strongly changes the spectrum of elementary excitations and bound states in contrast to the naive perturbation theory.

In the limit of massless quarks $m_u=0$ the left-hand side of Eq.~(\ref{sd-3u})
is equal to zero. The nonzero solution of Eq.~(\ref{sd-3u}) implies that
  there exists a mode with zero pion mass $M_{\pi}=0$
 in accordance with the Goldstone theorem.
 This means that the BS equation (\ref{bs-3u}), being the equation for the wave function of the Goldstone pion, coincides
with the the SD equation (\ref{sd-3u}) in the case of $m_u=M_{\pi}=0$.
Comparing the equations yields
 \bea\label{bs-4a}
 L^{\pi}_{1}(p)=\frac{M_u(p)}{F_\pi E_u(p)}=\frac{\cos2\upsilon_u(p)}{F_\pi},
 \eea
where the constant of the proportionality $F_\pi$ in Eq. (\ref{bs-4a}) is
called the  weak decay constant. In the more general case of massive quark $m_u\neq
M_{\pi}\neq0$, this constant is determined from the normalization condition
(\ref{nakap}) \bea
  \label{1fpi2}
  1 &=&\frac{4 N_c}{M_\pi} \int\limits_{ }^{ } \frac{d^3q}{(2\pi)^3} L_2 L_1=
 \frac{4 N_c}{M_\pi} \int\limits_{ }^{ } \frac{d^3q}{(2\pi)^3} L_2 \frac{\cos2\upsilon_u(q)}{F_\pi}
  \eea
with $N_c=3$.
In this case the wave function $L^{\pi}_{1}(p)$ is proportional
to the  Fourier component of the quark condensate
 \bea\label{1fpi2q}
 C_{\rm quark}=\sum_{n=1}^{n=N_c}\langle q_n(t,{\bf x})\overline{q}_n(t,{\bf y})\rangle=
 4N_c\int \frac{d^3 p}{(2\pi)^3}\frac{M_u(p)}{\sqrt{p^2+M^2_u(p)}}.
 \eea
Using Eqs.~(\ref{sd3}) and (\ref{bs-4a}), one can rewrite the definition of the quark condensate (\ref{1fpi2q}) in the form
\bea \label{gmor1}
 C_{\rm quark}&=&{4 N_c} \int\frac{d^3q}{(2\pi)^3}  \cos2\upsilon_u(q).
\eea

Let us assume that the representation for the wave function $L_1$ (\ref{bs-4a})  is still valid
for non-zero but small quark masses. Then
the subtraction of the BS equation (\ref{bs-3u}) from the SD one (\ref{sd-3u})
multiplied by the factor $1/F_\pi$ determines the second meson wave function
$L_2$
 \bea\label{bs-u}
 \frac{M_{\pi}}{2}L^{\pi}_{2}(p)&=&
\frac{m_u}{F_\pi}.
 \eea
The wave function $L^{\pi}_{2}(p)$ is independent of the
momentum in this aproximation. Substituting the equation $L_2=\text{const}=2m_u/(M_\pi
F_\pi)$ into the normalization condition (\ref{1fpi2}), and using
Eqs.~(\ref{bs-4a}) and (\ref{gmor1}), we arrive at the Gell-Mann-Oakes-Renner
relation \cite{gmor}
  \bea \label{gmor}
 M^2_{\pi}F^2_{\pi}&=&{2m_u} C_{\rm quark}\,.
\eea
 Our solutions including the GMOR relation (\ref{gmor}) differ from the accepted ones  \cite{puz,a4,a5,a6,a7,a10},
 where $\cos 2 {\upsilon}_{\mathrm{a}}(q)$
 is relaced by the sum of two Goldstone bosons, the pseudoscalar and the scalar one
   $[\cos 2 {\upsilon}_{\mathrm{a}}(q)+({\mathbf q}/q)\sin 2 {\upsilon}_{\mathrm{a}}(q)]$.
 This replacement can hardly been justified, because it is in contradiction with the
 Bethe-Salpeter equation (\ref{sca}) for scalar bound state with nonzero mass.

The coupled equations~(\ref{sd2}), (\ref{bs-3}), and~(\ref{bs-3-+}) contain the Goldstone mode that accompanies spontaneous breakdown of chiral symmetry. Thus, in  the framework of instantaneous interaction we prove the Goldstone theorem in the bilocal variant, and the GMOR relation directly results from the existence of the gluon and quark condensates. Strictly speaking, the postulate that the finiteness of the gluon and quark condensates are finite implies that QCD is the theory without ultraviolet  divergence. They can be removed by the Casimir type substraction \cite{sh-79} with the finite renormalization  \cite{Kaz-77}.

\subsection{New Hamiltonian interaction inspired by
the anomalous triangle diagram   with a pseudoscalar bound state}

  It was shown  \cite{pre-1a,pre-1} that the Habbard-Stratanovich linearization of the four fermion interaction
 leads to an effective action  for bound states in
 any gauge theory. We include here
 %the pion and pare-positronium $\eta_P$leads to
  an effective action describing the direct pion-positronium creation
 \be \label{s-11a}
  W_{eff}=\int  d^4x \left\{%\frac 1 2 \left[{\dot\eta_P}^2-M_P^2{\eta_P}^2-
%  \left(\partial_i\eta_P\right)^2
%  \right]
%   +
  \frac{\alpha}{\pi}\left(\frac{\pi_0}{F_\pi} + \frac{\eta_P}{F_P}\right) \dot A_i B_i+\frac{\dot A^2_i +B^2_i}{2}
   \right\}~,
  \ee
  where $\alpha=1/137$ is the QED coupling constant, and  $F_P$, contained in Eq. (\ref{t-1}),
  plays a role of the pion weak coupling parameter $F_\pi=92$ GeV.
  The first term $\dfrac{\alpha}{\pi} \left(\dfrac{\pi}{F_\pi}
   + \dfrac{\eta_P}{F_P}\right) \dot A_i B_i$ comes
  from the  triangle diagram (i.e., the anomalous term). This term
    describes the two $\gamma$ decay of   pseudoscalar bound states $P_{\rm
    bs}$.

     For each bound state one can obtain the corresponding two-photon anomalous creation
   cross section from Eq. (\ref{cp}). In the case of the process $\gamma+\gamma=P_{\rm bs}+P_{\rm bs}$
   we repeat Eq. (\ref{cp})
\bea\label{cp-1} \frac{d\sigma}{d\Omega} =\frac{\alpha^{4}E_\gamma^2}{\pi
128F^4_{P_{\rm bs}} }
\sqrt{1-{\left(\frac{2 m_e}{E_\gamma}\right)}^2}, %=\frac{\pi
%\alpha^{10}E_\gamma^2}{512 m_e^4} \sqrt{1-{\left(\frac{2
%m_e}{E_\gamma}\right)}^2}.
\eea
  where $P_{\rm bs}$  is the $F_\pi$ analogy.
In the case of the process $\gamma+\gamma=P_{\pi}+P_{\rm pos}$
   we obtain
\bea\label{cp-2} \frac{d\sigma}{d\Omega} =\frac{\alpha^{4}E_\gamma^2}{\pi 32 F^2_{P_\pi}F^2_{P_{\rm pos}}}
\sqrt{1-{\left(\frac{2 m_e}{E_\gamma}\right)}^2}.
\eea

 The Hamiltonian of this system is the sum of the energy of the free EM fields, the
    pseudoscalar Hamiltonians and their interactions
 \bea \label{s-12a}
  W_{eff}&=&\int dt d^3x \left\{
  E_i \dot A_i %+ P_\eta \dot \eta_P-{\cal H}_\eta
  -\frac{E_i^2+B_i^2}{2}-{\cal H}_{\rm int}\right\},\\%\label{s-13}
%  {\cal H}_\eta&=&\frac 1 2 \left({\dot\eta_P}^2-M_P^2{\eta_P}^2-(\partial_i{\eta_P})^2\right)\\
 \label{s-14a}
 {\cal H}_{\rm int }&=& \frac{\alpha}{\pi}\left(\frac{\pi_0}{F_\pi} + \frac{\eta_P}{\pi F_P}\right)  E_i B_i+
 \frac{\alpha^2}{2\pi^2} \left(\frac{\pi}{F_\pi} + \frac{\eta_P}{F_P}\right)^2 B_i^2~.
  \eea
This action contains the additional terms in comparison with the standard QED.
They leads to the additional  mass of the pseudoscalar bosons \cite{pavel-99}
 and the anomalous processes of the creation of the bound state pairs in the external magnetic field. The last term of the effective action (\ref{s-14a}) yields cross-sections of creation of both the two positronium atoms and  the pion and the positronium together.

The creation of two positronium atoms  is $\alpha^3$ times less than
the creation of the pion and the positronium together. In this case, one can speak about the pion catalysis of the positronium
  creation.

\section{Summary}

In this paper we  obtain the bound state functional by Poincar\'e-invariant
generalization of the FP path integral based on the Markov-Yukawa constraint
for description of both the spectrum equations and the S-matrix elements. The
axiomatic approach to gauge theories  presented here  allows us  to construct
the bound state functional in both QED and QCD on equal footing of the
Poincar\'e group representations.

It is shown that the Poincar\'e S-matrix, as compared with the Lorentz one,
contains
\begin{enumerate}
\item Creation of
 bound states inspired by the anomalous (triangle) diagram within the Hamiltonian approach.
\item This additional anomalous contribution includes the processes like $\gamma+\gamma \to Ps+ Ps$,
$\gamma+\gamma \to \pi_0+ Ps$, $\gamma+\gamma \to \pi_0+ \pi_0$ (where $Ps$ --
a pseudoscalar para-positronium).
\end{enumerate}

This raises the problem of  physical consequences of these additional processes.

The bound state generating functional  (\ref{fp-2d}),
where the time-axis is chosen as eigenvalue of the total momentum
operator of instantaneous bound states (\ref{3-2c}),
has a variety of properties. It
 describes spontaneous breakdown of chiral symmetry, the bilocal variant of the
 Goldstone theorem, and the direct derivation of the GMOR relation
 directly from
 the fact of existence of the finite gluon and quark condensates introduced by the normal
 ordering of the QCD action.
 The postulate of the finiteness of the gluon and quark condensates implies
 that both the QED and QCD can be considered on equal footing as
  the theory without ultraviolet  divergences. They can be  removed
 by the Casimir type substraction \cite{sh-79} with the finite renormalization  \cite{Kaz-77}.

\subsection*{Acknowledgments}
The authors would like to thank A.B. Arbuzov, B.M. Barbashov, A.A. Gusev,
A.V.~Efremov, O. V. Teryaev, R.N. Faustov, S.I. Vinitsky,
 and M.K. Volkov for useful discussions.
NSH  is grateful to the JINR Directorate for  hospitality.

\appendix
\section{Ladder approximation}\label{sect_AppA}

The generating functional (\ref{fp-3d}) can be presented by means of
the relativistic generalization of the
Hubbard-Stratonovich (HS) transformation \cite{pre-1a}
%The Hubbard-Stratonovich
% transformation is an exact mathematical transformation
\bea\label{eff-1a}
\exp[-ax^2/2]=[2\pi a]^{-1/2}\int_{-\infty}^{+\infty} dy\exp[-ixy -y^2/(2a)].
\eea
The basic idea of the HS transformation is to reformulate a system of
particles interacting through two-body potentials into a system of independent
particles interacting with a fluctuating field.
% invented by Russian physicist Ruslan L. Stratonovich and popularized by British physicist John Hubbard.
It is used to convert a particle theory into its respective field theory by
linearizing the density operator in the many-body interaction term of the
Hamiltonian and introducing a scalar auxiliary field  \cite{pre-1a}
 \bea     %\mathrm{a}
Z_{\psi}&=& \int d\psi
d\overline{\psi}e^{-\frac{i}{2}\left(\psi\overline{\psi},
{\cal K}\psi\overline{\psi}\right)-\left(\psi \overline{\psi},G^{-1}_{A^*}\right)+iS[J^*,\eta^*]}\\
\label{eff-1} &=&\int\left[\prod\limits_{x,y,a,b}^{} d {\cal
M}^{\mathrm{a}\mathrm{b}}(x,y)\right]\exp\{iW_{\rm eff}[{\cal M},A^*]
+i(\eta\overline{\eta},G_{{\cal M}})\}. \eea
 The effective action in Eq.
(\ref{eff-1}) can be decomposed in the form
\begin{eqnarray}
W_{eff}[{\cal M},A^*] = - {1 \over 2} N_{c} ( {\cal M}, {{\cal K}}^{-1} {\cal
M} ) + i N_{c} \sum_{n=1}^{\infty} {1 \over n} \Phi^{n} .
\end{eqnarray}
Here $ \Phi \equiv G_{A^*} {\cal M} , \Phi^{2} , \Phi^{3} $ etc. mean the
following expressions
\begin{eqnarray} \label{3-12a}
\Phi (x,y) \equiv G_{A^*} {\cal M} = \int d^{4}z G_{A^*} (x,z) {\cal M}(z,y), \nonumber \\
\Phi^{2} = \int d^{4} x d^{4}y \Phi(x,y) \Phi(y,x), \\
\Phi^{3} = \int d^{4} x d^{4}y d^{4}z \Phi(x,y) \Phi(y,z) \Phi(z,x) \,\,  ,
etc \nonumber
\end{eqnarray}

The first step to the semi-classical quantization of this construction
\cite{Eb-P-76} is the determination of its minimum of the effective action
\begin{eqnarray}\label{8}
N_{c}^{-1}{ \delta W_{eff} ({\cal M}) \over { \delta {\cal M}} } \equiv
 - {\cal K}^{-1} {\cal M} + {i \over { G_{A^*}^{-1} - {\cal M} } } = 0.
\end{eqnarray}
We denote the corresponding classical solution for the bilocal field by $
\Sigma (x-y) $. It depends only on the difference $x-y$ at $A^*=0$ because of
translation invariance of vacuum solutions.

The next step is the expansion of the effective action around the point of
minimum $ {\cal M} = \Sigma + {\cal M}^{\prime} $,
\begin{eqnarray}\label{9}
W_{eff} ( \Sigma + {\cal M}^{\prime} ) &= & W_{eff}^{(2)}+W_{int}; \nonumber \\
W_{eff}^{(2)} ({\cal M}^{\prime} ) &= & W_{Q}(\Sigma) + N_{c}[ - {1\over2}
{\cal M}^{\prime} {\cal K}^{-1} {\cal M}^{\prime}
 + { i \over 2} ( G_{\Sigma} {\cal M}^{\prime} )^{2} ] ; \nonumber \\
W_{int}=\sum_{n=3}^{\infty}W^{(n)}& = & i N_{c} \sum_{n=3}^{\infty} {1\over n}
( G_{\Sigma} {\cal M}^{\prime} )^{n}, \, \, \,~~~ ( G_{\Sigma} = (
G_{A^*}^{-1} - \Sigma)^{-1} ),
\end{eqnarray}
and the representation of the small fluctuations $ {\cal M}^{\prime} $  as a
sum (\ref{set-1})
\bea\label{set-1b} {\cal M}(x,y)&=&{\cal M}(z|X)\\\nonumber
&=&\sum\limits_H\int\frac{d^3{\cal
P}}{(2\pi)^{3}\sqrt{2\omega_H}}\int\frac{d^4qe^{iq\cdot
z}}{(2\pi)^4} \{ e^{i{\cal P}\cdot{X}}
\Gamma_H(q^{\bot}|{\cal P})a^+_H(\bm{{\cal P}})+e^{-i{\cal P}\cdot{X}}
\bar{\Gamma}_H(q^{\bot}|{\cal P})a^-_H(\bm{{\cal P}})\},
 \eea
%
% \bea\label{set} {\cal M}^{\prime}(z|X)=\sum\limits_H\int\frac{d^3
% {\cal P}}{(2\pi)^{3/2}\sqrt{2\omega_H}}\int\frac{d^4qe^{iq\cdot
% z}}{(2\pi)^4}\cdot\\\nonumber \{ e^{i\vec {\cal P}\vec{X}}
% \Gamma_H(q^{\bot}|{\cal P})a^+_H(\bm{{\cal P}})+e^{-i\vec {\cal P}\vec{X}}
% \bar{\Gamma}_H(q^{\bot}|-{\cal P})a^-_H(\bm{{\cal P}})\}
%  \eea
  over the complete set
of orthonormalized solutions $ \Gamma $, of the classical equation
\begin{eqnarray}\label{10}
%\left.
{ \delta^{2}W_{eff} ( \Sigma + {\cal M}^{\prime} ) \over { \delta {\cal
M}^{2}} }
%\right\vert_{ {\cal M}^{\prime} = 0}
\cdot \Gamma = 0
\end{eqnarray}
with a set of quantum numbers ($H$) including masses $M_H=\sqrt{{\cal
P}_\mu^2}$ and energies $\omega_H=\sqrt{{\bm{{\cal P}}}^2+M_H^2}$. The bound
state creation and annihilation operators obey the commutation relations
\begin{eqnarray} \label{comrel}
\biggl[ a^{-}_{H'}(\bm{{\cal P}}'), a^{+}_{H}(\bm{{\cal P}}) \biggr] =
\delta_{H'H} \delta^3 ( \bm{{\cal P}}' - \bm{{\cal P}} ) \,\,\, .
\end{eqnarray}

The  corresponding Green function takes the form \be\label{green} {\cal
G}(q^{\bot},p^{\bot}|{\cal P})=\sum\limits_H
\left\{\frac{\Gamma_H(q^{\bot}|{\cal P})\bar{\Gamma}_H(p^{\bot}|-{\cal P})}
{({\cal P}_0-\omega_H-i\varepsilon)2\omega_H}-
\frac{\bar\Gamma_H(p^{\bot}|{\cal P})){\Gamma}_H(p^{\bot}|-{\cal P})} {({\cal
P}_0-\omega_H-i\varepsilon)2\omega_H} \right\}~. \ee

To normalize vertex functions, we can use the ''free'' part of the effective action
(\ref{9}) for the quantum bilocal meson ${\cal M}^{\prime}$ with the
commutation relations~(\ref{comrel}). The substitution of the off - shell
$\sqrt{{\cal P}^2} \neq M_H$ decomposition (\ref{set-1}) into the "free" part of
effective action defines the reverse Green function of the bilocal field
${\cal G}({\cal P}_0)$
\begin{eqnarray}
W_{eff}^{(0)}[{\cal M}]=2\pi\delta({\cal P}_0-{\cal
P'}_0)\sum_H\int\frac{d^3{\cal P}}{\sqrt{2\omega_H}}a^{+}_{H}(\bm{{\cal P}})~
a^{-}_{H}(\bm{{\cal P}}){\cal G}^{-1}_H ({\cal P}_0)
\end{eqnarray}
where ${\cal G}^{-1}_H ({\cal P}_0) $ is the reverse Green function which can
be represented as a difference of two terms
\begin{eqnarray}
{\cal P}_H^{-1}({\cal P}_0) =  I( \sqrt{ {\cal P}^2 } ) -I(M_H^{\mathrm{a}\mathrm{b}}(\omega))
\end{eqnarray}
where $M_H^{ab}(\omega)$ is the eigenvalue of the equation for small
fluctuations~(\ref{10}) and
\begin{eqnarray}
I(\sqrt{{\cal P}^2}) &=&i N_c \int \frac{d^4 q}{(2\pi)^4} \times \nonumber \\
&&\mbox{tr} \biggl[ G_{\Sigma_{\mathrm{b}}}(q-\frac{{\cal P}}{2} ) \bar{\Gamma}_{\mathrm{a}\mathrm{b}}^H
(q^\perp | -{\cal P}) G_{\Sigma_{\mathrm{a}}}( q+ \frac{{\cal P}}{2} ) {\Gamma}_{\mathrm{a}\mathrm{b}}^H
(q^\perp | {\cal P}) \biggr]\nonumber
\end{eqnarray}
where \be\label{nak} \underline{G}_\Sigma(q)=\frac{1}{\not
q-\underline{\Sigma}(q^{\bot})},~~~~~~ \underline{\Sigma}(q) = \int d^{4}x
\Sigma(x) e^{iqx} \ee is the fermion Green function. The normalization
condition is defined by the formula
\begin{equation}
2 \omega = \frac{\partial{\cal G}^{-1}({\cal P}_0)}{\partial{\cal
P}_0}|_{{\cal P}_0=\omega({\cal P}_1)} = \frac{dM({\cal P}_0)}{d{\cal
P}_0}\frac{dI(M)}{dM}|_{{\cal P}_0=\omega}~.
\end{equation}
Finally, we get that solutions of equation (\ref{10}) satisfy the
normalization condition~\cite{nakanishi} \be\label{nakanishi} i N_c\frac{d}{d
{\cal P}_0}\int \frac{d^4q}{(2\pi)^{4}} \mbox{tr} \left[
\underline{G}_\Sigma(q-\frac{{\cal P}}{2})\bar{\Gamma}_H(q^{\bot}|-{\cal P})
\underline{G}_\Sigma(q+\frac{{\cal P}}{2}){\Gamma}_H(q^{\bot}|{\cal P})
\right] =2\omega_H. \ee

The achievement of the relativistic covariant constraint-shell quantization of
gauge theories is the description of both the spectrum of bound states and
their S-matrix elements.

It is convenient to write the relativistic-invariant matrix elements for the
action~~(\ref{9}) in terms of the field operator
$$
\Phi^{\prime}(x,y)=\int d^4x_1 G_{\Sigma}(x-x_1){\cal
M}'(x_1,y)=\Phi^{\prime}(z|X)
$$
Using the decomposition over the bound state quantum numbers $(H)$
  \bea\label{set1}
 {\Phi}^{\prime}(z|X)&=&\sum\limits_H\int\frac{d^3{\cal P}}{(2\pi)^{3/2}
 \sqrt{2\omega_H}}\int\frac{d^4q}{(2\pi)^4}\times\\
 &&\{ e^{i{\cal P}\cdot{X}} \Phi_H(q^{\bot}|{\cal P})a^+_H(\bm{{\cal P}})
 +e^{-i{\cal P}\cdot{X}} \bar{\Phi}_H(q^{\bot}|-{\cal P})a^-_H(\bm{{\cal P}})\}~,\nonumber
 \eea
where \be \Phi_{H(\mathrm{a}\mathrm{b})}(q^{\bot}|{\cal P})=G_{\Sigma a}(q+{\cal
P}/2)\Gamma_{H(\mathrm{a}\mathrm{b})}(q^{\bot}|{\cal P})~, \ee we can write the matrix elements
for the interaction $W^{(n)}$~(\ref{9}) between the vacuum and the n-bound
state \cite{yaf}
\begin{multline} \label{S-matrix}
\langle H_1{\cal P}_1, ...,H_n{\cal P}_n|iW^{(n)}|0\rangle =\\
=-i(2\pi)^4 \delta^4 \left( \sum\limits_{i=1 }^{n }{\cal P}_i\right)
\prod\limits_{j=1 }^{n } \left[\frac{1}{(2\pi)^32\omega_j}\right]^{1/2}
M^{(n)}({\cal P}_1,...,{\cal P}_n),
\end{multline}

\begin{multline} \label{S-matrix-1}
 M^{(n)}=\int \frac{id^4q}{(2\pi)^4 n} \sum\limits_{\{i_k\} }^{ }
 \Phi_{H_{i_{1}}}^{\mathrm{a}_1,\mathrm{a}_2}(q| {\cal P}_{i_1})\times \\
 \Phi_{H_{i_{2}}}^{\mathrm{a}_2,\mathrm{a}_3}(q-\frac{{\cal P}_{i_1}+{\cal P}_{i_2}}{2}| {\cal
 P}_{i_2}) \Phi_{H_{i_{3}}}^{\mathrm{a}_3,\mathrm{a}_4}\left(q-\frac{2{\cal P}_{i_2}+
 {\cal P}_{i_1}+{\cal P}_{i_3}}{2}| {\cal P}_{i_3}\right)\times \\
 ...\Phi_{H_{i_{n}}}^{\mathrm{a}_n,\mathrm{a}_1} \left(q-\frac{2({\cal P}_{i_2}+...+{\cal
 P}_{i_{n-1}})+{\cal P}_{i_1}+ {\cal P}_{i_n} }{2}| {\cal P}_{i_n}\right),
\end{multline}
where ($\{i_k\}$ denotes permutations over $i_k$).

Expressions % ~(\ref{set1a}),~
(\ref{green}),~(\ref{set1}),~(\ref{S-matrix}), and
~(\ref{S-matrix-1}) represent Feynman rules for the construction of a quantum
field theory with the action~(\ref{9}) in terms of bilocal fields.

%\section{Appendix C:}
\section{Bethe - Salpeter equation}\label{sect_AppB}

 Equations for the spectrum of the bound states (\ref{10})  can be
 rewritten in the form of the
 Bethe-Salpeter (BS) one~\cite{Salpeter}
 \begin{eqnarray}\label{bs}
 \Gamma = i {\cal K}(x,y) \int d^{4}z_{1} d^{4}z_{2}
 G_{\Sigma}(x-z_{1}) \Gamma(z_{1},z_{2}) G_{\Sigma}(z_{2}-y)~.
 \end{eqnarray}
In the momentum space with
 $$\underline{\Gamma}(q \vert {\cal P}) = \int d^{4}x d^{4}y
 e^{i{x+y \over2} {\cal P}} e^{i(x-y)q} \Gamma(x,y)
 $$
  %the Coulomb type kernel
   we obtain
   the following equation of
  the vertex function ( $ \underline
 {\Gamma} $ )
 \begin{eqnarray} \label{bs0}
 \underline{\Gamma}(k, {\cal P}) = i \int {d^{4}q \over (2\pi)^{4}}
 \underline {V} ( k^{\perp} - q^{\perp} ) \rlap/\ell \left[
 \underline{G}_{\Sigma}(q+{ {\cal P} \over 2 }) \Gamma(q \vert
 {\cal P} ) \underline{G}_{\Sigma}(q-{ {\cal P} \over 2 }) \right]
 \rlap/\ell
 \end{eqnarray}
 where $
 \underline {V} ( k^{\perp} ) $ means the Fourier transform of the
 potential, $ k^{\perp}_{\mu} = k_{\mu} - \ell_{\mu} ( k \cdot
 \ell) $ is the relative
 momentum transversal with respect to $ \ell_{\mu} $, and ${\cal P}_{\mu}$ is the total momentum.

 The quantity  $ \underline {\Gamma} $
 depends only on the transversal momentum
 \begin{eqnarray*}
 \underline{\Gamma}(k \vert {\cal P}) =
 \underline{\Gamma}(k^{\perp} \vert {\cal P}) ,
 \end{eqnarray*}
 because of the instantaneous form of the potential $
 \underline{V}(k^{\perp} )$ in any frame.

 We consider the Bethe - Salpeter equation~(\ref{bs})
 after integration over the longitudinal momentum
 $ q_{0} $. The vertex function takes the form
 \begin{eqnarray} \label{bs1}
 \Gamma_{\mathrm{a}\mathrm{b}}(k^{\perp} \vert {\cal P}) = \int { d^{3}q^{\perp}
 \over (2\pi)^{3} } \underline{V} ( k^{\perp}-q^{\perp} )
 \rlap/\ell \Psi_{\mathrm{a}\mathrm{b}}(q^{\perp}) \rlap/\ell,
 \end{eqnarray}
 where the bound state wave function $ \Psi_{\mathrm{a}\mathrm{b}} $ is given by
 \begin{eqnarray} \label{bs3}
 \Psi_{\mathrm{a}\mathrm{b}}(q^{\perp})= \rlap/\ell \left[ {
 \bar{\Lambda}_{(+)\mathrm{a}}(q^{\perp} \Gamma_{\mathrm{a}\mathrm{b}}(q^{\perp} \vert {\cal
 P} ) {\Lambda}_{(-)\mathrm{b}}(q^{\perp}) \over { E_{T} - \sqrt{ {\cal
 P}^{2} } + i \epsilon } } + { \bar{\Lambda}_{(-)\mathrm{a}}(q^{\perp}
 \Gamma_{\mathrm{a}\mathrm{b}}(q^{\perp} \vert {\cal P} ) {\Lambda}_{(+)\mathrm{b}}(q^{\perp})
 \over { E_{T} + \sqrt{ {\cal P}^{2} } - i \epsilon } }\right ]
 \rlap/\ell
 \end{eqnarray}
 $ E_{T} = E_{\mathrm{a}} + E_{\mathrm{b}} $ means
 the sum of one-particle energies of the two particles ($a$)
 and ($b$ ) defined by~(\ref{sd1}) %,~(\ref{sd2})
  and the notation~(\ref{ope1})
 \begin{eqnarray}  \label{ope2}
 \bar{\Lambda}_{(\pm)}(q^{\perp}) = S^{-1}(q^{\perp})
 \Lambda_{(\pm)}(0) S(q^{\perp}) = {\Lambda}_{(\pm)}( - q^{\perp})
 \end{eqnarray}
 has been introduced.

 Acting with the operators (\ref{ope1}) and (\ref{ope2}) on
 equation (\ref{bs1}) one gets
 the equations for the wave function $ \psi $ in an arbitrary moving
 reference frame
 \begin{eqnarray} \label{bs2}
 &&(E_{T}(k^{\perp}) \mp \sqrt { {\cal P}^{2}})
 \Lambda^{(\ell)}_{(\pm)\mathrm{a}}(k^{\perp}) \Psi_{\mathrm{a}\mathrm{b}}(k^{\perp})
 {\Lambda}^{(\ell)}_{(\mp)\mathrm{b}}( - k^{\perp}) = \nonumber \\  &=&
 \Lambda^{(\ell)}_{(\pm)a}(k^{\perp})  \int { d^{3}q^{\perp} \over
 (2\pi)^{3} } \underline{V} (k^{\perp}-q^{\perp})
 \Psi_{\mathrm{a}\mathrm{b}}(q^{\perp}) ] {\Lambda}^{(\ell)}_{(\mp)b}( - k^{\perp}) .
 \end{eqnarray}

 All these equations ~(\ref{bs1}) and ~(\ref{bs2})
 have been derived without any
 assumption about the smallness of the relative momentum $ \vert
 k^{\perp} \vert $ and for an arbitrary total momentum
 $$ {\cal
 P}_{\mu} = ( \sqrt { M_{A}^{2} + { {{\cal P}}}^{2} } , {\bm{{\cal P}}} \neq 0 )~.
 $$
 We expand the function $\Psi$ on the projection operators
 \be  \label{bs4}
 \Psi=\Psi_++\Psi_-,~~~~~\Psi_{\pm}=
 \Lambda^{(\ell)}_{\pm}\Psi
 \Lambda^{(\ell)}_{\mp}~.
 \ee
 According to Eq.~(\ref{bs3}), $\Psi$ satisfies the identities
 \be  \label{bs5}
 \Lambda^{(\ell)}_{+}\Psi\Lambda^{(\ell)}_{+}=
 \Lambda^{(\ell)}_{-}\Psi\Lambda^{(\ell)}_{-}\equiv 0~,
 \ee
 which permit the determination of an unambiguous expansion of $\Psi$
 in terms of the Lorentz structures:
 \be \label{bs6}
  \Psi_{\mathrm{a,b}\pm}=S^{-1}_{\mathrm{a}}\left\{ \gamma_5 L_{\mathrm{a,b}\pm}(q^{\bot})+
  (\gamma_{\mu}-\ell_{\mu}\not \ell)N^{\mu}_{\mathrm{a,b}\pm} \right\}
 \Lambda^{(\ell)}_{\mp}(0)S^{-1}_{\mathrm{b}}~,
 \ee
 where $L_{\pm}=L_1\pm L_2$, $N_{\pm}=N_1\pm N_2$. In the rest
 frame $\ell_{\mu}=(1,0,0,0)$ we get
 $$
 N^{\mu}=(0,N^i)~;
~~~~~~N^i(q)=\sum\limits_{a=1,2 }^{ }N_{\alpha}(q)e^i_{\alpha}(q)+
 \Sigma(q) \hat q^i~.
 $$
 The wave functions $L,N^{\alpha},\Sigma$  satisfy the following equations.

\begin{center}
\underline {1. Pseudoscalar particles.}
\end{center}

\begin{eqnarray}\label{pse}
 M_{L}  \stackrel{0}{L}_{2}({\bf p}) &=&
E  \stackrel{0}{L}_{1}({\bf p})
-
\int \frac{d^3 {q}}{(2\pi)^{3}}
V({\bf p}-{\bf q})
(  {\tt c}^{-}_{p} {\tt c}^{-}_{q}
- \xi {\tt s}^{-}_{p} {\tt s}^{-}_{q})
\stackrel{0}{L}_{1}({\bf q}) \,\,\, ;
\nonumber \\  \\
M_L  \stackrel{0}{L}_{1}({\bf p}) &=&
E \stackrel{0}{L}_{2}({\bf p})
-
\int \frac{d^3 {q}}{(2\pi)^{3}}
V({\bf p}-{\bf q})
(  {\tt c}^{+}_{p} {\tt c}^{+}_{q}
- \xi {\tt s}^{+}_{p} {\tt s}^{+}_{q})
\stackrel{0}{L}_{2} ({\bf q})     \,\,\, .
\nonumber
\end{eqnarray}
Here, in all equations, we use the following definitions
\bea\label{E-0}
E({\bf p})&=&E_{\mathrm{a}}({\bf p})+E_{\mathrm{b}}({\bf p})~,\\\label{E-1}
{\tt c}^{\pm}(p)&=&\cos[\upsilon_{\mathrm{a}}(p) \pm \upsilon_{\mathrm{b}}(p)]~,\\\label{E-2}
{\tt s}^{\pm}(p)&=&\sin[\upsilon_{\mathrm{a}}(p) \pm \upsilon_{\mathrm{b}}(p)]~,\\\label{E-3}
\xi &=& {\hat p}_{i} \cdot {\hat q}_{i} \,\,\, ,
\eea
where $E_{\mathrm{a}},E_{\mathrm{b}}$ are  one-particle energies and
$\upsilon_{\mathrm{a}}, \upsilon_{\mathrm{b}}$ are the Foldy-Wouthuysen angles of particles (a,b)
 given by Eqs.(\ref{sd2}) and (\ref{sd-2a}).

\begin{center}
\underline {2.Vector particles.}
\end{center}

\begin{eqnarray} \label{vecpar}
M_{\rm N} \stackrel{0}{N}_{2}{}^{\alpha} &=&
E \stackrel{0}{N}_{1}{}^{\alpha}  - \nonumber \\
&-&
\int \frac{d^3 {q}}{(2\pi)^{3}}
V({\bf p}-{\bf q})
\{
(  {\tt c}^{-}_{p} {\tt c}^{-}_{q}
{\underline \delta}^{\alpha\beta}
+ {\tt s}^{-}_{p} {\tt s}^{-}_{q}
( {\underline \delta}^{\alpha\beta} \xi -
\eta^{\alpha}{\underline \eta}^{\beta} ) )
\stackrel{0}{N}_{1}^{\beta}
+
(\eta^{\alpha} {\tt c}^{-}_{p} {\tt c}^{+}_{q})
\stackrel{0}{\Sigma}_{1} \}   \,\,\,  ; \nonumber \\  \\
M_{\rm N} \stackrel{0}{N}_{1}{}^{\alpha} &=&
E \stackrel{0}{N}_{2}{}^{\alpha}  -  \nonumber \\
&-&
\int \frac{d^3 {q}}{(2\pi)^{3}}
V({\bf p}-{\bf q})
\{
(  {\tt c}^{+}_{p} {\tt c}^{+}_{q}
{\underline \delta}^{\alpha\beta}
+ {\tt s}^{+}_{p} {\tt s}^{+}_{q}
( {\underline \delta}^{\alpha\beta} \xi -
\eta^{\alpha}{\underline \eta}^{\beta} ) )
\stackrel{0}{N}_{2}^{\beta}
+
(\eta^{\alpha} {\tt c}^{+}_{p} {\tt c}^{-}_{q})
\stackrel{0}{\Sigma}_{2}  \} \,\,\, .
\nonumber
\end{eqnarray}
\begin{eqnarray*}
\eta^{\alpha} = {\hat q}_{i} {\hat e}^{\alpha}_{i}(p) \,\,\, , \,\,\,
{\underline \eta}^{\alpha} = {\hat p}_{i} {\hat e}^{\alpha}_{i}(q) \,\,\, , \,\,\,
{\underline \delta}^{\alpha\beta}
= {\hat e}^{\alpha}_{i}(q){\hat e}^{\beta}_{i}(p) \,\,\, .
\end{eqnarray*}

\begin{center}
\underline {3. Scalar particles.}
\end{center}

\begin{eqnarray}\label{sca}
M_{\Sigma} \stackrel{0}{\Sigma}_{2} & = &
E \stackrel{0}{\Sigma}_{1}  -  \nonumber \\
&-&
\int \frac{d^3 {q}}{(2\pi)^{3}}
V({\bf p}-{\bf q})
\{
( \xi {\tt c}^{+}_{p} {\tt c}^{+}_{q} + {\tt s}^{+}_{p} {\tt s}^{+}_{q})
\stackrel{0}{\Sigma}_{1}
+
({\underline \eta}^{\beta}
{\tt c}^{-}_{p} {\tt c}^{+}_{q}) \stackrel{0}{N}_{1}{}^{\beta}  \} \,\,\, ; \nonumber \\  \\
M_{\Sigma} \stackrel{0}{\Sigma}_{1} & = &
E \stackrel{0}{\Sigma}_{2}   - \nonumber \\
&-&
\int \frac{d^3 {q}}{(2\pi)^{3}}
V({\bf p}-{\bf q})
\{
( \xi {\tt c}^{-}_{p} {\tt c}^{-}_{q} + {\tt s}^{-}_{p} {\tt s}^{-}_{q})
\stackrel{0}{\Sigma}_{2}
+                          ( {\underline \eta}^{\beta}
{\tt c}^{+}_{p} {\tt c}^{-}_{q}) \stackrel{0}{N_{2}}^{\beta} \} \,\,\, .  \nonumber
\end{eqnarray}

The normalization of these solutions is uniquely determined
by equation~(\ref{nakanishi})
\be\label{nakap}
\frac{2N_c}{M_L}\int \frac{d^3q}{(2\pi)^3}\left\{
L_1(q)L_2^*(q)+L_2(q)L_1^*(q)\right\}=1~,
\ee
\be\label{nakav}
\frac{2N_c}{M_N}\int  \frac{d^3q}{(2\pi)^3}\left\{
N^{\mu}_1(q)N^{\mu*}_2(q)+N^{\mu}_2(q)N^{\mu*}_1(q)\right\}=1~,
\ee
\be\label{nakas}
\frac{2N_c}{M_{\Sigma}}\int \frac{d^3q}{(2\pi)^3}\left\{
\Sigma_1(q)\Sigma_2^*(q)+\Sigma_2(q)\Sigma_1^*(q)\right\}=1~.
\ee

 If the atom is at rest ( $ {\cal P}_{\mu} = ( M_{A},0,0,0 ) $ )
 equation (\ref{bs2}) coincides with the Salpeter equation \cite{a15}. If
 one assumes that the current mass $ m^{0} $ is much larger than
 the relative momentum $\vert q^{\perp} \vert$, then the coupled
 equations ~(\ref{bs1}) and ~(\ref{bs2}) turn into the Schr\"o\-dinger
 equation. In the rest frame ( $ {\cal P}_{0} = M_{A} $) equation
 ~(\ref{sd1}) %and~(\ref{sd2})
  for a large mass  $ ( m^{0} / \vert q^{\perp} \vert
 \rightarrow \infty ) $ describes a nonrelativistic particle
 \begin{eqnarray*}
 E_{\mathrm{a}}( {\bf k} ) = \sqrt { ( m_{\mathrm{a}}^{0})^{2} + {\bf k}^{2} } \simeq
 m^{0}_{\mathrm{a}} + {1 \over 2} { {\bf k}^{2} \over m^{0}_{\mathrm{a}} }, \\
 \tan 2 \upsilon = { k \over m^{0} } \rightarrow 0 ; \,\, S({\bf
 k}) \simeq 1 ; \,\, \Lambda_{ (\pm)} \simeq { {1 \pm \gamma_{0} }
 \over 2 } .
 \end{eqnarray*}
 Then, in equation (\ref{bs2}) only the state with positive energy remains
 \begin{eqnarray*}\label{E-4}
  \Psi  \simeq \Psi_{(+)}= \Lambda_{(+)}\gamma_5\sqrt{4\mu}\psi_{\mathrm{Sch}}, \,\,\,
 \Lambda_{(-)} \psi \Lambda_{(+)} \simeq 0 ,
 \end{eqnarray*}
 where $ \mu = m_{\mathrm{a}} \cdot m_{\mathrm{b}} / ( m_{\mathrm{a}}+m_{\mathrm{b}}) $.
 And finally the Schr\"odinger equation results in
 \begin{eqnarray} \label{dinge}
 \left[ {1 \over 2\mu} {\bf k}^{-2} + ( m^{0}_{\mathrm{a}} + m^{0}_{\mathrm{b}} - M_{A} )
 \right] \psi_{\mathrm{Sch}}({\bf k}) = \int { d^3{q} \over (2\pi)^{3} }
 \underline{V} ({\bf k}-{\bf q}) \psi_{\mathrm{Sch}}({\bf q}),
 \end{eqnarray}
 with the normalization $\int d^3q|\psi_{\mathrm{Sch}}|^2/(2\pi)^3=1$.

 For an arbitrary total momentum $ {\cal P}_{\mu} $ equation~(\ref{dinge})
 takes the form
 \begin{eqnarray} \label{dinger}
 &&\left[- {1 \over 2\mu}
 (k_{\nu}^{\perp})^{-2} + ( m^{0}_{\mathrm{a}} + m^{0}_{\mathrm{b}} - \sqrt{ {\cal
 P}^{2}}  ) \right] \psi_{\mathrm{Sch}}( k^{\perp})
  = \int { d^{3} q^{\perp}
 \over (2\pi)^{3} } \underline{V} ( k^{\perp}- q^{\perp})
 \psi_{\mathrm{Sch}}( q^{\perp}),
 \end{eqnarray}
 and describes a
 relativistic atom with nonrelativistic relative momentum $ \vert
 k^{\perp} \vert \ll m^{0}_{a,b} $. In the framework of such a
 derivation of the Schr\"odinger equation it is sufficient to
 define the total coordinate as  $ X=(x+y)/2$,
 independently of the magnitude of the masses of the two particles
 forming an atom.
%
% In particular, the Coulomb interaction leads to a positronium
% at rest with the bilocal wave function (\ref{E-4})
% \bea\label{posi}
% \Psi_P^{\alpha\beta}(t|\vec z)&=&\eta_P(t)\left(\frac{1+\gamma_0}{2}
% \gamma_5\right)^{\alpha\beta}\underline{\psi}_{\mathrm{Sch}}(\vec z)
% \sqrt{\frac {m_e} 2}~,\\\label{po-sch}
% \underline{\psi}_{\mathrm{Sch}}(\vec z)&=&
% \int\limits_{ }^{ }\frac{d^3p}{(2\pi)^3}e^{(i\vec p \vec z)} \psi_{\mathrm{Sch}}(\vec p);
% \eea
% where $\underline{\psi}_{sch}(\vec z)$ is the Schr\"odinger
% normalizable wave function of the relative motion
% \be \label{relative}
% \left(-\frac 1{m_e}\frac {d^2}{d{\vec z}^2}
% -\frac {\alpha}{|\vec z|}\right)\underline{\psi}_{\mathrm{Sch}}(\vec z)=
% \epsilon\underline{\psi}_{\mathrm{Sch}}(\vec z);
%  ~~~~~~~~~~~\left(~\int d^3z{\parallel\underline{\psi}_{\mathrm{Sch}}(z)\parallel}^2=1\right)
% \ee
% $M_P=(2m_e-\epsilon)$ is the mass of a positronium,
% $\frac{1+\gamma_0} 2$ is the
% projection operator on the state with positive energies of an
% electron and positron.

\end{document}